\documentclass[12pt]{article}
\usepackage{amsmath}
\usepackage{graphicx}
\usepackage{natbib}
\usepackage{url} % not crucial - just used below for the URL 

\usepackage{amsthm}
\usepackage{amsmath}
\usepackage{amsfonts}
\usepackage{amssymb}
\usepackage{graphicx}
\usepackage{enumerate}
\usepackage{color}
\usepackage{array}
\usepackage{bbm}
\usepackage{multirow} 
\usepackage{rotating}
\usepackage{xr}
\usepackage{bm}
\usepackage{setspace}
\usepackage{pdflscape}
\usepackage{afterpage}
\usepackage{hyperref}
\usepackage{booktabs}

\usepackage{algorithm,algpseudocode}

\def\argmin{{\rm argmin}}

\begin{document}
\thispagestyle{empty}
\baselineskip=28pt
\vskip 5mm
\begin{center} {\Large{\bf Modern extreme value statistics \\ for Utopian extremes }} \\
 {{\bf EVA (2023) Conference Data Challenge: Team Yalla }}
\end{center}

\baselineskip=12pt
\vskip 5mm

\graphicspath{Figures}

\begin{center}
\large
Jordan Richards$^{1*}$, Noura Alotaibi$^2$, Daniela Cisneros$^2$, Yan Gong$^{3}$,\\Matheus B. Guerrero$^{4}$, Paolo Redondo$^2$,  Xuanjie Shao$^2$
\end{center}
\footnotetext[1]{
 \baselineskip=10pt {School of Mathematics, University of Edinburgh, UK}
}
\footnotetext[2]{
\baselineskip=10pt Statistics Program, Computer, Electrical and Mathematical Sciences and Engineering Division,\\ King Abdullah University of Science and Technology, Thuwal, Saudi Arabia.}
\footnotetext[3]{
\baselineskip=10pt {Harvard School of Public Health, Boston, Massachusetts, USA.}}

\footnotetext[4]{
\baselineskip=10pt Department of Mathematics, California State University Fullerton, California, {USA.}\\
$\;^*$Corresponding author: jordan.richards@ed.ac.uk
}

\baselineskip=17pt
\vskip 4mm
\centerline{\today}
\vskip 6mm

%%%%%%%%%%%%%%%%%%%%%%%%%%%%%%%%%%%%%%%%%%%%%%%%%%%%%%%%%%%%%%%%%%%%%%%%
\begin{center}
{\large{\bf Abstract}}\\
\end{center}
Capturing the extremal behaviour of data often requires bespoke marginal and dependence models which are grounded in rigorous asymptotic theory, and hence provide reliable extrapolation into the upper tails of the data-generating distribution. We present a toolbox of four methodological frameworks, motivated by modern extreme value theory, that can be used to accurately estimate extreme exceedance probabilities or the corresponding level in either a univariate or multivariate setting. Our frameworks were used to facilitate the winning contribution of Team Yalla to the EVA (2023) Conference Data Challenge, which was organised for the 13$^\text{th}$ International Conference on Extreme Value Analysis. This competition comprised seven teams competing across four separate sub-challenges, with each requiring the modelling of data simulated from known, yet highly complex, statistical distributions, and extrapolation far beyond the range of the available samples in order to predict probabilities of extreme events. Data were constructed to be representative of real environmental data, sampled from the fantasy country of ``Utopia''.

\baselineskip=16pt

\par\vfill\noindent
{\textbf{Keywords}: additive models; conditional extremes; neural Bayes estimation; non-parametric probability estimators; non-stationary extremal dependence; quantile regression}\\

\pagenumbering{arabic}
\baselineskip=24pt

\newpage

\section{Introduction}  %%%%%% --- New Section --- %%%

We are motivated by the EVA (2023) Conference Data Challenge organised for the 13$^\text{th}$ International Conference on Extreme Value Analysis, with full details provided in the editorial by \cite{editorial}. In this paper, we detail the four modelling frameworks used by ``Team Yalla'' to win the aforementioned challenge, which comprised four sub-challenges that each required prediction of exceedance probabilities or quantiles for data simulated from highly complex statistical models. These frameworks combined classical EVA methods with modern modelling techniques, including additive models \citep{chavez2005generalized,youngman2019generalized}, deep learning-based inference \citep{sainsbury2022fast,richards2023likelihood}, non-stationary conditional extremal dependence models \citep{heffernan2004conditional,Winter2016}, and non-parametric probability estimators \citep{krupskii2019nonparametric}. Whilst the data considered in this work are simulated, the data were constructed to be representative of real observations of an environmental process (sampled from the fantasy country of ``Utopia''), and so exhibit realistic characteristics, such as sparsity, non-stationarity, and missingness. Moreover, as the true values of the challenge predictands are known, our predictions can be easily validated. Hence, we expect our proposed frameworks to perform well in practice, with real data. {The code used for implementing our models is available at \url{https://github.com/matheusguerrero/yalla}. }

The novelty of this work is threefold: i) we propose an amortised neural Bayes estimator for univariate quantiles; ii) we generalise the non-parametric multivariate exceedance probability estimator of \cite{krupskii2019nonparametric}; and iii) we illustrate the efficacy of the extreme value regression models proposed by \cite{Winter2016} and \cite{youngman2019generalized} when applied to simulated data. The remainder of the paper is organised as follows: Section~\ref{sec:methodology} describes the methodology we adopt to address the data challenge, with Sections~\ref{sec:POT}--\ref{sec:NBE} and \ref{sec:CEM}--\ref{sec:nonpar} focusing on univariate and multivariate extremes, respectively. In Section~\ref{sec:res}, we apply our proposed methodology to the data. Section~\ref{Sec:conclusion} provides a concise conclusion and suggests avenues for further work.

\section{Methodology}\label{sec:methodology} % --- %%%
%%%%%%%%%%%%%%%%%%%%%%%%%%%%%%%%%%%%%%%%%%%%%%%%%%%%%%
%%% --- New Section --- %%%%%% --- New Section --- %%%
%%%%%%%%%%%%%%%%%%%%%%%%%%%%%%%%%%%%%%%%%%%%%%%%%%%%%%

{In this section, we present the methodological details of our approaches. Throughout, we adopt the notation $Y$ and $\mathbf{Y}:=(Y_1,\dots,Y_d)^{\prime}$ to denote a random response variable and random $d$-vector of response variables, respectively. Covariates are denoted by $\mathbf{X}:=(X_1,\dots,X_l)^{\prime}\in\mathbb{R}^l$ for $l\in\mathbb{N}$. Where appropriate, we use the subscript $t\in\{1,\dots,n\}$ to denote temporal replicates of the response (i.e., $Y_t$ or $\mathbf{Y}_t:=(Y_{1,t},\dots,Y_{d,t})^{\prime}$) or covariates (i.e., $\mathbf{X}_t:=(X_{1,t},\dots,X_{l,t})^{\prime}$) and lowercase notation to denote observations. The values of $l$, $d$, and sample size ${n}$ differ throughout the paper. Sub-challenges C1/C2 and C3/C4 \citep[see][]{editorial} concern univariate and multivariate modelling, respectively. In the univariate setting with $d=1$ and $n=21000$, covariate} vector $\mathbf{X}$ comprises $l=8$ variables: wind direction, wind speed, atmosphere, season, and four unnamed variables (denoted $V_1,\dots,V_4)$. For C3 and C4, we have $d=3$ and $d=50$, respectively, and $n=21000$ and $n=10000$, respectively, with the response vector $\mathbf{Y}$ known to have standard Gumbel margins. The three response variables for C3, $(Y_1,Y_2,Y_3)$, are accompanied by the atmosphere and season covariates described above; hence, $l=2$ for C3. No covariates accompany the 50 response variables that comprise the data for C4 (i.e., $l=0$). 

 Sections~\ref{sec:POT} and \ref{sec:NBE} detail methodology for estimating univariate extreme quantiles, whilst Sections~\ref{sec:CEM}-\ref{sec:nonpar} concern methodology for modelling extremal dependence.  Section~\ref{sec:POT} describes peaks-over-threshold modelling using the generalised Pareto distribution and generalised additive models, which we used to address sub-challenge C1. Section~\ref{sec:NBE} describes a likelihood-free neural Bayes estimator for point estimation of the extreme quantiles for sub-challenge C2. Section~\ref{sec:CEM} describes a non-stationary model for conditional extremal dependence (sub-challenge C3), whilst Section~\ref{sec:EDM} details a bivariate extremal dependence measure. Section~\ref{sec:nonpar} concludes with details of a non-parametric estimator for tail probabilities, used in sub-challenge C4. 

%%%%%%%%%%%%%%%%%%%%%%%%%%%%%%%%%%%%%%%%%%%%%%%%%%%%%%%%%%%%
%%% --- New Subsection --- %%%%%% --- New Subsection --- %%%
%%%%%%%%%%%%%%%%%%%%%%%%%%%%%%%%%%%%%%%%%%%%%%%%%%%%%%%%%%%%
\subsection{Peaks-over-threshold models}\label{sec:POT}  %%%
%%%%%%%%%%%%%%%%%%%%%%%%%%%%%%%%%%%%%%%%%%%%%%%%%%%%%%%%%%%%
%%% --- New Subsection --- %%%%%% --- New Subsection --- %%%
%%%%%%%%%%%%%%%%%%%%%%%%%%%%%%%%%%%%%%%%%%%%%%%%%%%%%%%%%%%%
{Sub-challenge C1 required the prediction of a $50\%$ confidence interval for the $q-$quantile of $Y\mid (\mathbf{X}=\mathbf{x}^*_j)$ for 100 test covariate sets $\{\mathbf{x}^*_j:j=1,\dots,100\},$ and for $q=0.9999$ corresponding to an extreme conditional quantile. To this end, we adopt a peaks-over-threshold regression model.} The upper tails of the distribution of a random variable can be modelled in this framework using the generalised Pareto distribution (GPD), see, for example, \cite{davison1990models}. For a random variable $Y$, we assume that there exists some high threshold $u$ such that the distribution of exceedances $(Y-u)\mid( Y > u)$ can be characterised by the GPD, denoted GPD$(\sigma_u,\xi)$, $\sigma_u>0,\xi\in\mathbb{R}$, with distribution function 
\begin{align}
H(y) = 
\begin{cases}
1-(1+\xi y /\sigma_u)^{-1/\xi}, \quad & \xi\neq0, \\
1-\exp(-y/\sigma_u), \quad & \xi=0,
\end{cases}
\end{align}
where $y \geq 0$ for $\xi \geq0 $ and $0 \leq y \leq -\sigma_u/\xi$ for $\xi < 0$.

For regression, we model the conditional distribution $(Y-u(\mathbf{x}))\mid{( Y > u(\mathbf{x}),\mathbf{X}=\mathbf{x})}$ as $\mbox{GPD}(\sigma_u(\mathbf{x}),\xi(\mathbf{x}))$, where the exceedance threshold and GPD scale and shape parameters are functions of covariates. We follow \cite{chavez2005generalized} and \cite{youngman2019generalized} and use a generalised additive model (GAM) representation for the distribution, with $\sigma_u(\mathbf{x})$ and $\xi(\mathbf{x})$ modelled via a basis of splines. The threshold $u(\mathbf{x})$ is taken to be the intermediate $\lambda$-quantile of $Y\mid(\mathbf{X}=\mathbf{x})$ for $\lambda < 0.9999$ and this is modelled using additive quantile regression \citep{fasiolo2021fast}.

%%%%%%%%%%%%%%%%%%%%%%%%%%%%%%%%%%%%%%%%%%%%%%%%%%%%%%%%%%%%
%%% --- New Subsection --- %%%%%% --- New Subsection --- %%%
%%%%%%%%%%%%%%%%%%%%%%%%%%%%%%%%%%%%%%%%%%%%%%%%%%%%%%%%%%%%
\subsection{Neural point estimation}\label{sec:NBE} %--- %%%
%%%%%%%%%%%%%%%%%%%%%%%%%%%%%%%%%%%%%%%%%%%%%%%%%%%%%%%%%%%%
%%% --- New Subsection --- %%%%%% --- New Subsection --- %%%
%%%%%%%%%%%%%%%%%%%%%%%%%%%%%%%%%%%%%%%%%%%%%%%%%%%%%%%%%%%%
{Sub-challenge C2 required estimation of the $q$-quantile of $Y$ for ${q=1-(6\times 10^4)^{-1}}$, i.e., estimating $\theta$ such that $\Pr\{Y>\theta \}=1-q$. However, inference for $\theta$ should seek to minimise the conservative asymmetric loss function
\begin{equation}\label{eq:asym_loss}
L(\theta,\hat{\theta})=
\begin{cases}
0.9(0.99\theta-\hat{\theta}),\quad&\text{if } 0.99\theta > \hat{\theta},\\
0,\quad &\text{if } \vert\theta-\hat{\theta}\vert \leq 0.01\theta,\\
0.1(\hat{\theta}-1.01\theta),\quad &\text{if } 1.01\theta < \hat{\theta},
\end{cases}
\end{equation}
where $\hat{\theta}$ denotes estimates of $\theta$ and $\vert\cdot\vert$ denotes the absolute value; this loss function is illustrated in Figure~\ref{fig:NBE_res}. The loss function in Eq.~\eqref{eq:asym_loss} provides a larger penalty for under-estimates of $\theta$, relative to over-estimates, to encourage conservative estimates of the quantile. We construct a conservative estimator for extreme quantiles using neural networks. }

Neural point estimators, that is, neural networks that are trained to map input data to parameter point estimates, have shown recent success as a likelihood-free inference approach for statistical models. Although they have been typically used for inference with classical spatial processes \citep[see, e.g.,][]{zammit2020deep,gerber2021fast} and spatial extremal processes \citep[see, e.g.,][]{lenzi2021neural,lenzi2023towards,sainsbury2023neural,sainsbury2022fast, richards2023likelihood}, they can be exploited in a univariate setting. For example, \cite{rai2023fast} use a neural point estimator to make inference with the univariate generalised extreme value distribution \citep[see][]{coles2001introduction}. We construct a neural point estimator to perform extreme single quantile estimation for a random variable $Z$. In particular, we follow \cite{sainsbury2022fast} and construct a neural Bayes estimator (NBE).

Define a set of univariate probability distributions $\mathcal{P}$ on a sample space, taken to be $\mathbb{R}$, which are parameterised by a parameter ${\theta}\in\mathbb{R}$ such that $\mathcal{P}\equiv \{P_{{\theta}}:{\theta}\in\Theta\}$, where $\Theta$ is the parameter space; then $\mathcal{P}$ defines a parametric statistical model \citep[see][]{mccullagh2002statistical}. Denote $\mathbf{Z}\equiv({Z}_1,\dots,{Z}_n)^{\prime}$ as $n$ mutually independent realisations {of the random variable $Z$} from $P_{{\theta}} \in \mathcal{P}$. A point estimator $\hat{{\theta}}(\cdot)$ for model $\mathcal{P}$ is any mapping from $\mathbb{R}^n$ to $\Theta$, and the output of such an estimator, for a given ${\theta}$ and $\mathbf{Z}$, can be assessed using a non-negative loss function $L({\theta},\hat{{\theta}}(\mathbf{Z}))$. The risk of this point estimator, evaluated at ${\theta}$, $R({\theta},\hat{{\theta}}(\cdot)),$ is the loss $L(\cdot,\cdot)$ averaged over all possible realisations of $\mathbf{Z}$, that is,
\begin{equation}
\label{eq:risk}
R({\theta},\hat{{\theta}}(\cdot))\equiv \int_{\mathbb{R}^n}L({\theta},\hat{{\theta}}(\mathbf{z}))f(\mathbf{z}\mid{\theta})\mathop{}\!\mathrm{d}\mathbf{z},
\end{equation}
where $f(\mathbf{z}\mid{\theta})$ is the density function of the data. We define the Bayes risk $r_\pi(\cdot)$ as the weighted average of Eq.~\eqref{eq:risk} over all ${\theta}\in\Theta$, with respect to some prior measure $\pi(\cdot)$, as
\begin{equation}
\label{eq:Bayesrisk}
r_\pi(\hat{{\theta}}(\cdot))\equiv\int_\Theta R({\theta},\hat{{\theta}}(\cdot))\mathop{}\!\mathrm{d}\pi({\theta}).
\end{equation}

If the estimator $\hat{{\theta}}(\cdot)$ minimises Eq.~\eqref{eq:Bayesrisk}, we term it a \textit{Bayes estimator} with respect to the loss $L(\cdot,\cdot)$ and prior measure $\pi(\cdot)$. Note that in the context of the prediction task, the parameter ${\theta}$ is taken to be the $q$-quantile of the distribution(s) $P_\theta$ and the loss function is $L(\theta,\hat{\theta})$ in Eq.~\eqref{eq:asym_loss} (illustrated in Figure~\ref{fig:NBE_res}). Details on the construction of the prior measure $\pi(\cdot)$ and the statistical model $\mathcal{P}$ will follow.

A neural Bayes estimator (NBE) is a neural network designed to approximately minimise Eq.~\eqref{eq:Bayesrisk}. \cite{sainsbury2022fast} construct NBEs by leveraging the DeepSets neural network architecture \citep{zaheer2017deep}. Consider functions $\boldsymbol{\psi}:\mathbb{R}\mapsto \mathbb{R}^Q$ and ${\phi}:\mathbb{R}^Q\mapsto\mathbb{R}$, and a permutation-invariant set function $\mathbf{a}:(\mathbb{R}^Q)^n\mapsto\mathbb{R}^Q$, where the $j$-th component of $\mathbf{a}$, $a_j(\cdot),$ returns the element-wise average over its input set for $j=1,\dots,Q$. We represent ${\phi}(\cdot)$ and $\boldsymbol{\psi}(\cdot)$ as neural networks, and collect in $\boldsymbol{\gamma}\equiv (\boldsymbol{\gamma}_\phi^{\prime},\boldsymbol{\gamma}_{\boldsymbol{\psi}}^{\prime})^{\prime}$ their estimable ``weights'' and ``biases''. Our NBE is of the form
\begin{equation}
\label{eq:NE}
\hat{{\theta}}(\mathbf{Y}; \boldsymbol{\gamma})={\phi}(\mathbf{T}(\mathbf{Z}; \boldsymbol{\gamma}_{{\boldsymbol{\psi}}}); \boldsymbol{\gamma}_{{\phi}}),\quad\text{ with }\quad\mathbf{T}(\mathbf{Z}; \boldsymbol{\gamma}_{\boldsymbol{\psi}})=\mathbf{a}(\{\boldsymbol{\psi}(Z_t; \boldsymbol{\gamma}_{\boldsymbol{\psi}}):t=1,\dots,n\}).
\end{equation}
We use a densely-connected neural network to model both ${\phi}(\cdot)$ and $\boldsymbol{\psi}(\cdot)$. The NBE is built by obtaining neural network weights $\boldsymbol{\gamma}^*$ that minimise the Bayes risk in the estimator space spanned by $\hat{{\theta}}(\cdot,\boldsymbol{\gamma})$. As we cannot directly evaluate Eq.~\eqref{eq:Bayesrisk}, it is approximated using Monte Carlo methods. {For a set of $K$ parameter values $\{\theta^{(k)}:k=1,\dots,K\}$  drawn from the prior $\pi(\cdot)$, we simulate, for each $k$, a set of $n$ mutually independent realisations $\mathbf{z}^{(k)}$ from $P_{{\theta}}$. The Bayes risk in Eq.~\eqref{eq:Bayesrisk} is then approximated by
\begin{equation}
\label{Eq:risk}
\hat{r}_\pi(\hat{{\theta}}(\cdot;\boldsymbol{\gamma}))=\frac{1}{K}\sum_{k=1}^KL({\theta}^{(k)},\hat{{\theta}}(\mathbf{z}^{(k)};\boldsymbol{\gamma}))\approx {r}_\pi(\hat{{\theta}}(\cdot;\boldsymbol{\gamma})).
\end{equation}}
We obtain estimates $\boldsymbol{\gamma}^*=\argmin_{\boldsymbol{\gamma}}\hat{r}_\pi(\hat{{\theta}}(\cdot;\boldsymbol{\gamma}))$ using the package \texttt{NeuralEstimators} \citep{sainsbury2022fast} in $\texttt{Julia}$ \citep{bezanson2017julia}. \\
{Specifying a prior measure on the $q$-quantile, $\theta \in \Theta$, for a random variable $Z$ is nontrivial, as the mapping from the model $\mathcal{P}$ to the parameter space $\Theta$ is not guaranteed to be surjective; for a fixed level $q$, multiple probability distributions can have the same value of $\theta$, and so simulating $Z$ conditional on $\theta$ is not necessarily feasible. In this case, $P_\theta$ would define a set of distributions with equal $q$-quantile, rather than a single probability distribution (as in \cite{sainsbury2022fast}). Thus, instead of placing a prior directly on $\theta$,
we assume that $P_\theta$ is determined by some hyper-parameters (for simplicity, we omit the dependency of $P_\theta$ on hyper-parameters from our notation). We then construct a general prior measure for these hyper-parameters, which consequently induces a prior measure on $\theta$. As we are interested in $\theta$ when $q$ is close to one, our choice of the class of feasible models $\mathcal{P}$ is motivated by extreme value theory, and we exploit the univariate peaks-over-threshold models described in Section~\ref{sec:POT}. 
}

{\cite{editorial} note that our data $\{Y_t:t=1,\dots,n\},$ from which we wish to infer $\theta$, are non-stationary over time. We reflect this property in our construction for the prior measure on the hyper-parameters of $P_\theta$: for $t=1,\dots,n$, let $(Y_t-u_t)\mid(Y_t > u_t)\sim \mbox{GPD}(\sigma_t,\xi_t)$ and let $Y_t\mid{(Y_t \leq u_t)}\sim F^\leq_t(\cdot)$ for threshold $u_t\in\mathbb{R}$, scale $\sigma_t>0,$ and shape $\xi_t \in \mathbb{R}$, and where $\Pr\{Y_t \leq u_t\} = \lambda$ for $\lambda \in [0,1]$. For simplicity, we hereafter treat $\lambda$ and the distribution of non-exceedances $F^\leq_t(\cdot)$ as fixed. After placing a suitable prior measure on the hyper-parameters $\{(u_t,\sigma_t,\xi_t): t=1,\dots,n\}$ and specifying $\lambda$ and  $F^\leq_t(\cdot)$ (see, e.g., Section~\ref{sec:res:uni}), we can simulate data $\{y^*_t:t=1,\dots,n^*\}$ from the model above and store this in a vector $\mathbf{z}^*$ with the index $t$ removed from each entry. In this way, we consider $\mathbf{z}^*$ as $n^*$ independent draws from the distribution of $Z$, unconditional on $t$, as we have marginalised out the effect of this covariate \citep[see, e.g.,][]{rohrbeck2018extreme}. Note that $n^*$ need not satisfy $n^*=n$, where $n$ is the sample size. } 

{The prior measure on the hyper-parameter set  $\{(u_t,\sigma_t,\xi_t): t=1,\dots,n\}$ induces a prior on $\theta$, which is the $q$-quantile of $Z$.} As there is no closed-form expression for $\theta$, we compute it using Monte Carlo methods. That is, {we set $n^*$ large and derive $\theta$ empirically from realisations $\mathbf{z}^*$. A single entry to the training data for our NBE then consists of the pair $(\mathbf{z},\theta)$, where $\mathbf{z}$ is a sub-sample from $\mathbf{z}^*$ of length $n$; this procedure is repeated for a total of $K$ entries. Note that, we could instead construct a prior on $\theta$ using a stationary GPD model, i.e., with $n=1$, but our approach produces a more diffuse prior on $\theta$ by increasing the prior support for the hyper-parameters of ${P}_\theta$. It also exploits knowledge about the data generating distribution, which produces more realistic models for $Z$.  } 

 {When choosing larger values of $\lambda$, fewer new observations are generated for training of the NBE; a bigger proportion of the training data comprise resamples from the observations. 
Smaller values of $\lambda$ will produce a more diffuse prior on the distributional models for $Z$ (particularly with regards to the upper-tail behaviour of $Z$) and, hence, the $q$-quantile of $Z$. A larger variety in the training data is also likely to increase the reliability of our estimator when generalising to unseen data. However, when $\lambda$ is too small, the threshold $u_t$ may be too low to safely assume that $(Y_t-u_t)\mid (Y_t > u_t)$ follows a GPD. In this case, our estimator would not benefit from being trained on well-specified models for $Y_t$.  Hence, we advocate taking $\lambda$ as low as possible whilst still obtaining reasonable fits for the non-stationary GPD model, even if this value is not optimal. In our application, we take $\lambda=0.6$ but note that this is lower than the optimal $\lambda$ (in terms of providing the best fit for the model described in Section~\ref{sec:POT}).}
%%%%%%%%%%%%%%%%%%%%%%%%%%%%%%%%%%%%%%%%%%%%%%%%%%%%%%%%%%%%
%%% --- New Subsection --- %%%%%% --- New Subsection --- %%%
%%%%%%%%%%%%%%%%%%%%%%%%%%%%%%%%%%%%%%%%%%%%%%%%%%%%%%%%%%%%
\subsection{Conditional {extremes} models}
\label{sec:CEM}
%%%%%%%%%%%%%%%%%%%%%%%%%%%%%%%%%%%%%%%%%%%%%%%%%%%%%%%%%%%%
%%% --- New Subsection --- %%%%%% --- New Subsection --- %%%
%%%%%%%%%%%%%%%%%%%%%%%%%%%%%%%%%%%%%%%%%%%%%%%%%%%%%%%%%%%%
{Sub-challenge C3 required estimation of
\begin{align}
\label{Eq:C3}
 p_1 &:= \Pr(Y_1 > 6, Y_2 > 6, Y_3 > 6),\\\nonumber
 p_2 &:= \Pr(Y_1 > 7, Y_2 > 7, Y_3 < -\log{(\log{2})} ),
\end{align}
where $ -\log{(\log{2})}$ is the median of the standard Gumbel distribution. To estimate these extreme exceedance probabilities, we construct a non-stationary extremal dependence model for random vectors.} Proposed by \cite{heffernan2004conditional} and later generalised by \cite{Heffernan2007}, the conditional extremes framework models the behaviour of a random vector, conditional on one of its components being extreme. We adopt this model as its inference is typically less computationally demanding than that of other models for multivariate extremal dependence \citep{huser2022advances}. Additionally, it is capable of capturing both asymptotic dependence and asymptotic independence in a parsimonious manner. To accommodate non-stationarity with respect to covariates in the extremal dependence structure of our data, we utilise the extension of the \cite{heffernan2004conditional} model proposed by \cite{Winter2016}. This extension represents the dependence parameters as a linear function (subject to some non-linear link transformation) of the covariates.

Let the vector $\mathbf{Y}_t\equiv(Y_{1,t},Y_{2,t},Y_{3,t})^{\prime}$ for $t\in\{1,\dots,n\}$ have standard Laplace margins \citep{keef2013estimation}. It is noteworthy that our original data are known to possess standard Gumbel marginals; therefore, we transform these to have standard Laplace margins. Then, denote by $\mathbf{Y}_{-i,t}$ as the vector $\mathbf{Y}_t$ with its $i$-th component removed. Note that all vector operations hereafter are taken component-wise. \cite{Winter2016} assume that there exist vectors of coefficients $\boldsymbol{\alpha}_{-i,t}:=\{\alpha_{j\mid i,t}: j\in(1,2,3) \setminus i\}\in[-1,1]^{2}$ and $\boldsymbol{\beta}_{-i,t}:=\{\beta_{j\mid i,t}: j\in(1,2,3)\setminus i\}\in[0,1)^{2}$ such that, for $\mathbf{z}\in\mathbb{R}^{2}$ and $y>0$,
\begin{equation}
\label{Eq:heff}
\Pr\left\{\frac{\mathbf{Y}_{-i,t}-\boldsymbol{\alpha}_{-i,t}Y_{i,t}}{Y_{i,t}^{\boldsymbol{\beta}_{-i,t}}}\leq \mathbf{z}, Y_{i,t}-u > y \;\;\bigg\vert\;\; Y_{i,t} > u\right\}\rightarrow G_{-i,t}(\mathbf{z})\exp(-y),
\end{equation}
as $u\rightarrow \infty$ with $G_{-i,t}(\cdot)$ a non-degenerate bivariate distribution function. The values of the dependence parameters $\boldsymbol{\alpha}_{-i,t}$ and $\boldsymbol{\beta}_{-i,t}$ determine the strength and class of extremal dependence exhibited between $Y_{i,t}$ and the corresponding component of $\mathbf{Y}_{-i,t}$; for details, see \cite{heffernan2004conditional}. We allow these parameters to vary with covariates $\mathbf{x}_t$ by letting
\begin{equation}
    \label{eq:C3-paramform}
    \tanh^{-1}(\boldsymbol{\alpha}_{-i,t}):=\boldsymbol{\alpha}^{(0)}_{-i}+\boldsymbol{\alpha}^{(1)}_{-i}\mathbf{x}_t,\quad \text{logit}(\boldsymbol{\beta}_{-i,t}):=\boldsymbol{\beta}^{(0)}_{-i}+\boldsymbol{\beta}^{(1)}_{-i}\mathbf{x}_t,
\end{equation}
with coefficients $\boldsymbol{\alpha}^{(0)}_{-i}$,$\;\boldsymbol{\beta}^{(0)}_{-i}\in\mathbb{R}^2$ and $\boldsymbol{\alpha}^{(1)}_{-i}$,$\;\boldsymbol{\beta}^{(1)}_{-i}\in\mathbb{R}^{2\times l},$ where $l$ is the number of covariates. Note that the $\tanh(\cdot)$ and $\text{logit}(\cdot)$ link functions are used to ensure that the parameter values are constrained to their correct ranges.

Modelling follows under the assumption that the limit in Eq.~\eqref{Eq:heff} holds in equality for all $Y_{i,t} > u$ for some sufficiently high threshold $u>0$. In this case, rearranging  Eq.~\eqref{Eq:heff} provides the model
\begin{equation}
\label{Eq:heff2}
\mathbf{Y}_{-i,t}=\boldsymbol{\alpha}_{-i,t}Y_{i,t}+Y_{i,t}^{\boldsymbol{\beta}_{-i,t}}\mathbf{Z}_{i,t}\; \big\vert\; (Y_{i,t} > u),
\end{equation}
where the residual random vector $\mathbf{Z}_{i,t}:=\{Z_{j \mid i,t}: j\in(1,2,3)\setminus i\}\sim G_{-i,t}$ is independent of $Y_{i,t}$. For inference, we make the working assumption that $G_{-i,t}(\cdot)$ does not depend on time $t$ and follows a bivariate standard Gaussian copula with correlation $\rho_i\in (-1,1)$ and delta-Laplace margins \citep[see, e.g., ][]{shooter2021basin}. We hereafter drop the subscript $t$ from the notation. A random variable that follows the delta-Laplace distribution with location, scale, and shape parameters $\mu\in\mathbb{R},\;\sigma>0,$ and $\delta >0$, respectively, has density function $f(z)=\delta (2k\sigma\Gamma(\delta^{-1}))^{-1}\exp\{-(|z-\mu|/(k\sigma))^{\delta}\}$ for $k^2=\Gamma(\delta^{-1})/\Gamma(3\delta^{-1}),$ where $\Gamma(\cdot)$ denotes the standard gamma function. Note that when $\delta=1$ and $\delta=2$, we have the Laplace and Gaussian densities, respectively. 

Inference proceeds via maximum likelihood estimation. The model is fitted separately for each conditioning variable $i=1,2,3$. For each $i$, we have eight parameters in $\boldsymbol{\alpha}^{(0)}_{-i}$,$\;\boldsymbol{\alpha}^{(1)}_{-i}$,$\;\boldsymbol{\beta}^{(0)}_{-i}$,$\;\boldsymbol{\beta}^{(1)}_{-i},$ as well as the seven parameters that characterise $G_{-i}$, that is, the correlation $\rho_i$ and the three marginal parameters for each component of   $\mathbf{Z}_{i,t}$. After estimation of the parameters, we no longer require the working Gaussian copula assumption for $\mathbf{Z}_{i,t}$. We instead use observations ${\mathbf{y}}_{-i,t}$ and ${{y}}_{i,t}$ to derive the empirical residual vector\[
\tilde{\mathbf{z}}_{i,t}:=({\mathbf{y}}_{-i,t}-\hat{\boldsymbol{\alpha}}_{-i,t}{y}_{i,t})/{{{y}}_{i,t}}^{\hat{\boldsymbol{\beta}}_{-i,t}},
\]
where $\hat{\boldsymbol{\alpha}}_{-i,t}$ and $\hat{\boldsymbol{\beta}}_{-i,t}$ denote estimates of ${\boldsymbol{\alpha}_{-i,t}}$ and ${\boldsymbol{\beta}_{-i,t}}$, respectively. Assuming independence across time, we use the empirical residuals to provide an empirical estimate $\tilde{G}_{-i}(\cdot)$ of $G_{-i}(\cdot)$.

We estimate the necessary exceedance probabilities, $p_1$ and $p_2$ in Eq.~\eqref{Eq:C3}, via the following Monte-Carlo procedure. We first note that, for fixed $t\in\{1,\dots,n\}$, we require realisations of $\mathbf{Y}_t$, i.e., unconditional on an exceedance. We follow, for example, \cite{richards2022modelling, richards2023joint} and obtain these realisations by drawing a realisation from 
\begin{equation}
\label{Eq:heff3}
\mathbf{Y}_{t} \;\bigg\vert \left(\max_{i=1,2,3}Y_{i,t} > u\right),
\end{equation}
with probability 
\begin{equation}
\label{Eq:heff4}
\Pr\left\{\max_{i=1,2,3}Y_{i,t} > u\right\},
\end{equation}
and, otherwise, drawing a realisation of $\mathbf{Y}_{t} \mid (\max_{i=1,2,3}Y_{i,t} < u)$. As realisations of the latter are unlikely to significantly impact estimates of $p_1$ and $p_2$, we draw them empirically \citep{richards2022modelling}. In practice, we also replace {the} probability in Eq.~\eqref{Eq:heff4} with an empirical estimate. In addition, as both $\mathbf{Y}_{t} \mid (\max_{i=1,2,3}Y_{i,t} < u)$ and Eq.~\eqref{Eq:heff4} depend on $t$, we estimate them empirically by assuming stationarity over time.

To simulate from Eq.~\eqref{Eq:heff3}, we must first draw realisations of the conditional exceedance model, $\mathbf{Y}_{-i,t} \mid (Y_{i,t} > u)$ in Eq.~\eqref{Eq:heff2}, using Algorithm~\ref{sim-algo}. 
\begin{algorithm}[!htb]
\caption{Simulating from Eq.~\eqref{Eq:heff2}.}
\label{sim-algo}
For time $t$ and conditioning index $i\in\{1,2,3\}$:
\begin{enumerate}
\item Simulate $E \sim \mbox{Exp}(1)$ and set $y_{i,t} = u + E$.
\item \label{sim-algo:2} Draw a residual vector ${\tilde{\mathbf{z}}}\sim\tilde{G}_{-i}(\cdot)$.
\item Set $\mathbf{y}_{-i,t}=\hat{\boldsymbol{\alpha}}_{-i,t}y_{i,t}+y_{i,t}^{\hat{\boldsymbol{\beta}}_{-i,t}}\tilde{\mathbf{z}}$.
\end{enumerate}
\end{algorithm}
We can then combine realisations from the three separate conditional exceedance models, i.e., $\mathbf{Y}_{-i,t} \mid (Y_{i,t} > u)$ for each $i=1,2,3$, into a single realisation of Eq.~\eqref{Eq:heff3} using importance sampling \citep{wadsworth2022higher}. Whilst this can be achieved for a single fixed value of $t$, we note that $p_1$ and $p_2$ in Eq.~\eqref{Eq:C3} do not depend on the time $t$. Hence, we treat $t$ as random during simulation and average over all times $t\in\{1,\dots,n\}$ to produce approximate realisations of $\mathbf{Y}$ with $t$ marginalised out. The full simulation algorithm is detailed in Algorithm~\ref{sim-algo2}. Note that we back-transform from standard Laplace to standard Gumbel margins after simulation, as the original data has the latter.
\begin{algorithm}[!bht]
\caption{Simulating $\{(Y_{1,t},Y_{2,t},Y_{3,t}):t\in\{1,\dots,n\}\}$}
\label{sim-algo2}
\begin{enumerate}
\item For $j = 1,\dots, N^{\prime}$ with $N^{\prime} > N$: \begin{enumerate}
\item Draw a conditioning index $i_j\in\{1,2,3\}$  with equal probability.
\item Draw a time $t_j\in\{1,\dots,n\}$  with equal probability.
\item Simulate $\mathbf{y}^{(j)}$ using Algorithm~\ref{sim-algo} with time $t_j$ and conditioning index $i_j$.
\end{enumerate}
\item Assign each simulated vector $\mathbf{y}^{(j)}:=(y_1^{(j)},y_2^{(j)},y_3^{(j)})^{\prime}$  an importance weight of
\[
\left\{\sum_{i=1}^3\mathbbm{1}\{{y}^{(j)}_{i}>u\}\right\}^{-1},
\]
for $j=1,\dots,N^{\prime}$, and sub-sample $N$ realisations from the collection with probabilities proportional to these weights.
\item With probability $n^{-1}\sum^n_{t=1}\mathbbm{1}\{\max_{i=1,2,3}{y}_{i,t} <u\}$, re-sample (with equal probability) from \[\left\{{\mathbf{y}}_t: t\in\{1,\dots,n\},\max\limits_{i=1,2,3}{y}_{i,t} <u\right\}.\]
\item Back-transform sample onto standard Gumbel margins.
\end{enumerate}
\end{algorithm}

%%%%%%%%%%%%%%%%%%%%%%%%%%%%%%%%%%%%%%%%%%%%%%%%%%%%%%%%%%%%
%%% --- New Subsection --- %%%%%% --- New Subsection --- %%%
%%%%%%%%%%%%%%%%%%%%%%%%%%%%%%%%%%%%%%%%%%%%%%%%%%%%%%%%%%%%
\subsection{Extremal dependence measure}\label{sec:EDM}  %%%
%%%%%%%%%%%%%%%%%%%%%%%%%%%%%%%%%%%%%%%%%%%%%%%%%%%%%%%%%%%%
%%% --- New Subsection --- %%%%%% --- New Subsection --- %%%
%%%%%%%%%%%%%%%%%%%%%%%%%%%%%%%%%%%%%%%%%%%%%%%%%%%%%%%%%%%%

A number of pairwise measures have been proposed to quantify the strength of extremal dependence between random variables $(Y_1,Y_2)$, see, for example, \cite{heffernan2000directory}. We adopt the \textit{extremal dependence measure} (EDM) proposed by \cite{resnick2004extremal} and \cite{larsson2012extremal}, which has been extended to a multivariate setting by \cite{cooley2019decompositions}. Let $(Y_1,Y_2)\in[0,\infty)^2$ be a regularly-varying random vector with index $\alpha >0$; for full details on regular variation, see \cite{resnick2007heavy}. For some symmetric norm $\|\cdot\|$ on $[0,\infty)^2$, define the transformation $(R,\boldsymbol{\Omega}):=(\|(Y_1,Y_2)\|,(Y_1,Y_2)/\|(Y_1,Y_2)\|)$. Then, there exists a sequence $b_n\rightarrow \infty$ and constant $c>0$ such that
\[
n\Pr\{(b_n^{-1}R, \boldsymbol{\Omega})\in\cdot)\xrightarrow{v} c\nu_\alpha \times H,
\]
where $\xrightarrow{v}$ denotes vague convergence, $\nu_\alpha$ is a measure on $(0,\infty]$ such that $\nu_\alpha((y,\infty])= y^{-\alpha}$ and the angular measure $H$ is defined on the unit circle $\mathcal{S}_+=\{\mathbf{y} \in [0,\infty)^2\setminus \{\mathbf{0}\}: \|\mathbf{y}\|=1\}$. The extremal dependence measure \citep{larsson2012extremal} between $Y_1$ and $Y_2$ is
\[
\mbox{EDM}(Y_1,Y_2)=\int_{\mathcal{S}_+}\omega_1\omega_2\mathop{}\!\mathrm{d}H(\boldsymbol{\omega})=\lim_{y\rightarrow \infty}\mathbb{E}\left(\Omega_1\Omega_2 \mid R > y\right).
\]
Note that $\mbox{EDM}(Y_1,Y_2)\in[0,1]$ quantifies the strength of extreme dependence between $Y_1$ and $Y_2$. If $\mbox{EDM}(Y_1,Y_2) = 0$, then $H$ concentrates all mass along the axes and hence $Y_1$ and $Y_2$ are asymptotically independent \citep{resnick2007heavy}. Conversely,  $\mbox{EDM}(Y_1,Y_2)$ is maximal if and only if $(Y_1,Y_2)$ has full asymptotic dependence, or $H$ places all mass on the diagonal. \cite{larsson2012extremal} propose the empirical estimator 
\begin{equation}\label{EmpEDM}
\widehat{\operatorname{EDM}}\left(Y_1, Y_2\right)=\frac{1}{N_n} \sum_{t=1}^n \frac{y_{1 , t}}{r_t} \frac{y_{2, t}}{r_t} \mathbbm{1}{\left\{r_t>u\right\}},
\end{equation}
where $\bm{y}_t:=\left(y_{1 , t}, y_{2 ,t}\right)$, $t=1, \ldots, n$, is an i.i.d sample from $\left(Y_1, Y_2\right)$, $r_t=\left\|\bm{y}_t\right\|$, and $N_n=\sum_{t=1}^n \mathbbm{1}{\left\{r_t>u\right\}}$ is the number of exceedances of $\{r_t:t=1,\dots,n\}$ above some high threshold $u>0$. In Section~\ref{sec:result-dep}, we use pairwise EDM estimates to investigate the extremal dependence structure of a high-dimensional random vector and then decompose it into subvectors of strongly tail-dependent variables.

%%%%%%%%%%%%%%%%%%%%%%%%%%%%%%%%%%%%%%%%%%%%%%%%%%%%%%%%%%%%
%%% --- New Subsection --- %%%%%% --- New Subsection --- %%%
%%%%%%%%%%%%%%%%%%%%%%%%%%%%%%%%%%%%%%%%%%%%%%%%%%%%%%%%%%%%
\subsection{Non-parametric tail probability estimation}
\label{sec:nonpar}
%%%%%%%%%%%%%%%%%%%%%%%%%%%%%%%%%%%%%%%%%%%%%%%%%%%%%%%%%%%%
%%% --- New Subsection --- %%%%%% --- New Subsection --- %%%
%%%%%%%%%%%%%%%%%%%%%%%%%%%%%%%%%%%%%%%%%%%%%%%%%%%%%%%%%%%%

{
Sub-challenge C4 requires estimation of two exceedance probabilities of the form
\begin{align}
\label{Eq:C4}
 p_3 &:= \Pr(Y_1 >s_1, \dots, Y_{50} >s_1),\nonumber\\
 p_4 &:= \Pr(Y_1 > s_1,\dots,Y_{25}>s_1,Y_{26} > s_2,\dots,Y_{50}>s_2),
\end{align}
where $s_j,j=1,2,$ is the $(1-\phi_j)$-quantile of the standard Gumbel distribution with $\phi_1=1/300$ and $\phi_2=12\phi_1$. Hence, $p_3$ corresponds to an exceedance probability with all components of $\mathbf{Y}$ being equally extreme, i.e., concurrently exceeding the same marginal quantile, whilst, for $p_4$, the first 25 components of $\mathbf{Y}$ exceed a higher quantile, i.e., are more extreme, than the latter 25 components. }Thus, we seek an estimator for exceedance probabilities of high-dimensional multivariate random vectors $\mathbf{Y}\in\mathbb{R}^d$. While scalable parametric models for high-dimensional multivariate extremes do exist, as exemplified by \cite{engelke2021sparse} and \cite{lederer2023extremes}, they often impose restrictive assumptions about extremal dependence in $\mathbf{Y}$, particularly as the dimension $d$ grows large. Examples include regular or hidden regular variation \citep{resnick2002hidden}. As our goal is point estimation of exceedance probabilities, rather than a full characterisation of the joint upper tail of $\mathbf{Y}$, we choose instead to use a non-parametric estimator for exceedance probabilities that makes few assumptions on the joint upper tails. In particular, we adopt the non-parametric multivariate tail probability estimator proposed by \cite{krupskii2019nonparametric}. 

Denote by $F_i(\cdot)$ the distribution function\footnote{In our application to the prediction challenge data, $F_i(\cdot)$ is known to be the standard Gumbel distribution function for all $i=1,\dots,d$; see \cite{editorial}.} of $Y_i$ and let $U_i=F_i(Y_i)$ for $i=1,\dots,d$. Then, for $\phi\in(0,1)$, let
\begin{equation}
\label{eq:pq}
p(\phi) := \Pr\{U_1 > 1-\phi, \ldots, U_{d} > 1-\phi\}=\Pr\{U_{\text{max}}  < \phi\},
\end{equation}
where $U_{\text{max}}=\max(U_1,\dots,U_d)$. Denote by $C:[0,1]^d\mapsto\mathbb{R}$ the copula associated with $\mathbf{Y}$ and by $\bar{C}$ the corresponding survival copula, such that $p(\phi)=\bar{C}(1-\phi,\dots,1-\phi)$. \cite{krupskii2019nonparametric} construct estimators of $p(\phi)$ for small $\phi>0$ by making assumptions about the joint tail decay of $C$. In particular, they assume that $C$ (and $\bar{C}$) has continuous partial derivatives and, as $\phi\downarrow 0$, that
\begin{align}
\label{eq:pq2}
p(\phi) &=\Pr\{U_{\text{max}}<\phi\}= \lambda_1 \phi+\lambda_2 \phi^{1/\eta}  \ell(\phi)+o\{\phi^{1/\eta} \ell(\phi)\},\nonumber\\
\frac{\mathrm{d} p(\phi)}{\mathrm{d}\phi} &= \lambda_1+\lambda_2 \phi^{1/\eta-1}\ell(\phi)/\eta+o\{\phi^{1/\eta-1} \ell(\phi)\},
\end{align}
for $\lambda_1,\lambda_2 \geq 0,$ and $\eta<1$, and where $\ell(\cdot)$ is a slowly-varying function at zero such that, for all $s>0$,  $\ell(s\phi)/\ell(\phi)\rightarrow 0$ as $\phi\downarrow 0$; these regularity assumptions hold for a number of popular parametric copulas \citep[see][]{krupskii2019nonparametric}. We note that, when $d=2$, $\eta$ corresponds to the coefficient of tail dependence proposed by \cite{ledford1996statistics}; the extension to $d$-dimensions is described by \cite{eastoe2012modelling}.

Under the model in Eq.~\eqref{eq:pq2}, the parameters $\lambda_1,\lambda_2,$ and $\eta$ can be estimated by defining a small threshold $\phi_*>\phi$, and considering $U_{\text{max}} \mid U_{\text{max}} <\phi_*$. Then, as $\phi\downarrow 0$,
\begin{equation}
\label{eq:pq3}
\Pr\{U_{\text{max}} < \phi \mid U_{\text{max}} < \phi_*\}=p(\phi)/p(\phi_*)\sim k_1 \phi + k_2 \phi^{1/\eta},
\end{equation}
where $k_1=\lambda_1/(\lambda_1\phi_*+\lambda \phi_*^{1/\eta})$ and $k_2=(1-k_1\phi_*)/\phi_*^{1/\eta}$. By fixing $\phi_*$ and assuming equality between the left and right-hand sides of Eq.~\eqref{eq:pq3}, parameter estimates $\hat{k}_1$ and $\hat{\eta}_1$ can be computed using maximum likelihood methods. Then, our estimate of $p(\phi)$ is
\begin{equation}
\label{eq:pq4}
\widehat{p(\phi)}:=\{\hat{k}_1\phi+(1-\hat{k}_1\phi_*)\phi^{1/\hat{\eta}}\}\widehat{p(\phi_*)},
\end{equation}
where $\widehat{p(\phi_*)}$ is the empirical estimate of $p(\phi_*)$.

The estimator in Eq.~\eqref{eq:pq4} can be directly used to estimate $p_3$ in Eq.~\eqref{Eq:C4} by setting $\phi=\phi_1=1/300$. However, it cannot be used to estimate $p_4$, as the formulation of the joint survival probability in Eq.~\eqref{eq:pq} does not account for different levels of marginal tail decay for each component of $\mathbf{Y}$. In fact, we require an adaptation of $p(\phi)$ such that each component $U_i,i=1,\dots,d,$ exceeds a scaled value of the form $1-c_i\phi$ for $0<c_i<1/\phi$. Setting $c_i=1$ for $i=1,\dots,25$ and $c_i=12$, otherwise, yields $p_3$ in Eq.~\eqref{Eq:C4}. We construct such an estimator by considering, for $\mathbf{c}:=(c_1,\dots,c_d)^{\prime}$, the probability 
\begin{equation}
\label{eq:pq5}
p(\phi, \mathbf{c}) := \Pr\{U_1 > 1-c_1\phi, \ldots, U_{d} > 1-c_d\phi\}=\Pr\{U_{\text{max}}(\mathbf{c})  < \phi\},
\end{equation}
for weighted maxima $U_{\text{max}}(\mathbf{c}):=\max\{(1-U_1)/c_1,\dots,(1-U_d)/c_d\}$; note that when $\mathbf{c}=(1,\dots,1)'$, we have equivalence between Eq.~\eqref{eq:pq} and Eq.~\eqref{eq:pq5}. An estimator of the form in Eq.~\eqref{eq:pq5} was alluded to by \cite{krupskii2019nonparametric}; however, they did not provide theoretical results for its asymptotic behaviour. We choose to estimate $p(\phi, \mathbf{c})$ by assuming that, as $\phi\downarrow 0$, $\Pr\{U_{\text{max}}(\mathbf{c})  < \phi\}$ has the same parametric form as $p(\phi)$ in Eq.~\eqref{eq:pq2}. Inference for $p(\phi, \mathbf{c})$ then follows in a similar manner as for $p(\phi)$, only with samples of $U_{\text{max}}(\mathbf{c}) \mid (U_{\text{max}}(\mathbf{c}) > \phi_*)$ (rather than $U_{\text{max}} \mid (U_{\text{max}} > \phi_*)$) used for inference.

%%%%%%%%%%%%%%%%%%%%%%%%%%%%%%%%%%%%%%%%%%%%%%%%%%%%%%%%%%%%
%%% --- End of Section --- %%%%%% --- End of Section --- %%%
%%%%%%%%%%%%%%%%%%%%%%%%%%%%%%%%%%%%%%%%%%%%%%%%%%%%%%%%%%%%
%%%%%%%%%%%%%%%%%%%%%%%%%%%%%%%%%%%%%%%%%%%%%%%%%%%%%%%%%%%%
%%% --- End of Section --- %%%%%% --- End of Section --- %%%
%%%%%%%%%%%%%%%%%%%%%%%%%%%%%%%%%%%%%%%%%%%%%%%%%%%%%%%%%%%%

%%%%%%%%%%%%%%%%%%%%%%%%%%%%%%%%%%%%%%%%%%%%%%%%%%%%%%
%%% --- New Section --- %%%%%% --- New Section --- %%%
%%%%%%%%%%%%%%%%%%%%%%%%%%%%%%%%%%%%%%%%%%%%%%%%%%%%%%
\section{Results}\label{sec:res} % New Section --- %%%
%%%%%%%%%%%%%%%%%%%%%%%%%%%%%%%%%%%%%%%%%%%%%%%%%%%%%%
%%% --- New Section --- %%%%%% --- New Section --- %%%
%%%%%%%%%%%%%%%%%%%%%%%%%%%%%%%%%%%%%%%%%%%%%%%%%%%%%%

%%%%%%%%%%%%%%%%%%%%%%%%%%%%%%%%%%%%%%%%%%%%%%%%%%%%%%%%%%%%
%%% --- New Subsection --- %%%%%% --- New Subsection --- %%%
%%%%%%%%%%%%%%%%%%%%%%%%%%%%%%%%%%%%%%%%%%%%%%%%%%%%%%%%%%%%
\subsection{Univariate quantile estimation}
\label{sec:res:uni}
%%%%%%%%%%%%%%%%%%%%%%%%%%%%%%%%%%%%%%%%%%%%%%%%%%%%%%%%%%%%
%%% --- New Subsection --- %%%%%% --- New Subsection --- %%%
%%%%%%%%%%%%%%%%%%%%%%%%%%%%%%%%%%%%%%%%%%%%%%%%%%%%%%%%%%%%

We now describe estimation of {the} univariate conditional quantiles required for the data prediction challenge \citep{editorial}. To estimate the conditional $q-$quantile of $Y \mid (\mathbf{X}=\mathbf{x})$ for $q=0.9999$, we fit the GPD-GAM model described in Section~\ref{sec:POT}. The structure of $u(\mathbf{x})$, $\sigma_u(\mathbf{x})$, and $\xi(\mathbf{x})$ are optimised, for a fixed value of $\lambda$, by minimising the model's BIC for different additive combinations of linear and smooth functions of the covariates. {All smooth terms are centered at zero and represented as univariate thin-plate splines, with their degrees of freedom optimised automatically using the default penalisation options available in the \texttt{R} package \texttt{evgam} \citep{evgam}. Missing covariate values are imputed to the marginal mean of the observed values and missingness is treated as a factor variable. That is, smooth terms in a model output one of two values, depending on whether or not the input covariate is missing.}

We optimise $\lambda$ by using the threshold selection scheme proposed by \cite{varty2021inference} and extended by \cite{murphy2023automated}. We first follow \cite{heffernan2001extreme} and use the quantile and GPD-GAM model estimates to transform all data onto standard exponential margins. Then, we define a grid of $n$ equally spaced probabilities, $\{0<\lambda_1<...<\lambda_{n}<1$\}, with $\lambda_n=1-(\lambda_2-\lambda_1)$. Let $F_E(\cdot)$ be the standard exponential distribution function and $\lambda_* \in (0,1)$ be a pre-specified cut-off threshold. For all $i=1,\dots,n$, we define a vector of weights $\mathbf{w}=(w_1,\dots,w_n)^{\prime}$ with components $w_i=F_{E}^{-1}(\lambda_i)/\sum^n_{j=1}F_{E}^{-1}(\lambda_j)$ for $\lambda_i>\lambda_*$ and zero, otherwise. We then define the tail-weighted standardised mean absolute deviance (twsMAD) by $(1/n)\sum^n_{i=1}w_i\lvert\tilde{\theta}(\lambda_i)-F_E^{-1}(\lambda_i)\rvert,$ where $\tilde{\theta}(\lambda_i)$ denotes the empirical $\lambda_i$-quantile of the standardised data. We set $\lambda_*=0.99$ and evaluate the twsMAD for a range of candidate $\lambda$ values that subceed $\lambda_*$. We then choose the optimal value which minimises this metric.

\begin{figure}[!hbt]
\centering
\includegraphics[width=0.5\linewidth]{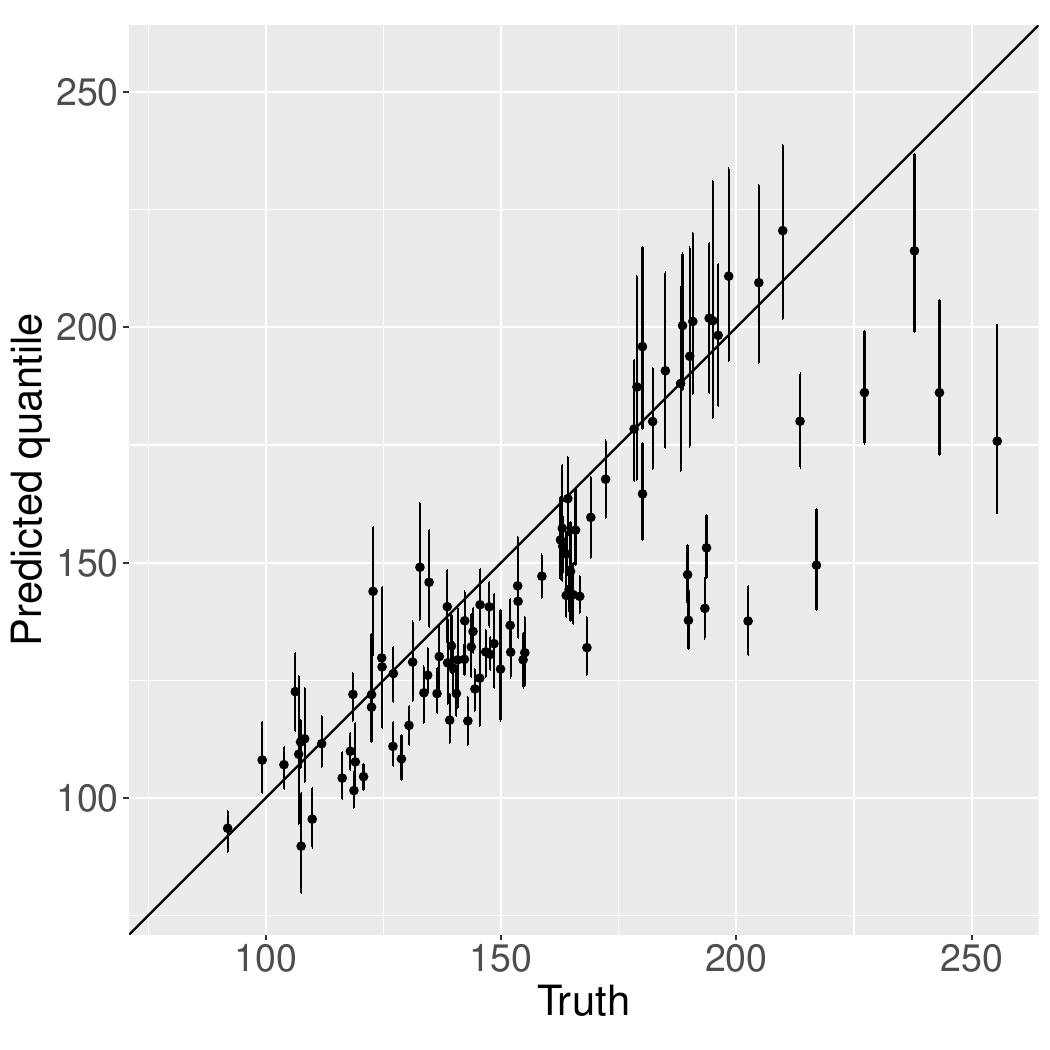} 
\caption{Predicted conditional quantiles against true values for the 100 test covariate sets for sub-challenge C1. Vertical error bars denote the $50\%$ confidence interval, with the black points denoting the median over bootstrap samples. }
\label{fig:C1}
\end{figure}

Using the BIC and twsMAD, our final model uses $\lambda=0.972$ and, for the covariates, we include: season and wind speed as linear terms in $u(\mathbf{x})$, $\sigma_u(\mathbf{x}),$ and $\xi(\mathbf{x})$; $V_2$, $V_3$, $V_4$, and wind direction as smooth terms in $u(\mathbf{x})$ and $\sigma_u(\mathbf{x})$. The median and the $50\%$ confidence intervals for the test conditional $q$-quantiles (see Section~\ref{sec:POT}) are estimated using the empirical quantiles over $2500$ non-parametric bootstrap samples, and presented in Figure~\ref{fig:C1}. The true values of the quantiles are provided by \cite{editorial}. We observe good predictions of the conditional quantiles, with our estimates close to the true values. Our framework attains a $38\%$ coverage rate, which is a slight underestimation of the normal $50\%$.

To estimate the required extreme $q$-quantile $\theta$ of $Y$, we use the NBE framework detailed in Section~\ref{sec:NBE} with $n=21000$ independent replicates. Neural networks are trained using the Adaptive Moment Estimation (Adam) algorithm \citep{kingma2014adam} with a mini-batch size of 32 and, in order to improve numerical stability, input data are standardised prior to training by subtracting and dividing by the mean and standard deviation (evaluated over the entire training data set), respectively. We use $K=350,000$ and $K/5$ parameter values for training and validation, respectively, with $n^*=4\times 10^7$ replicates used to estimate the theoretical $q$-quantiles ($\theta$; see Section~\ref{sec:NBE}). When producing training data, we use an empirical prior distribution for the hyper-parameter set $\{(u_t,\sigma_t,\xi_t): t=1,\dots,n\}$. To construct this prior, we estimate the GPD-GAM model, as described above but with $\lambda=0.6$ (see Section~\ref{sec:NBE} for details), across 750 bootstrap samples. Draws from the empirical prior on $\{(u_t,\sigma_t,\xi_t):t=1,\dots,n\}$ then correspond to a sample from all bootstrap parameter estimates. To increase the variety of models (and hence, quantiles) used to train the estimator, we permute the values of $u_t$, $\sigma_t,$ and $\xi_t$ across $\{(u_t,\sigma_t,\xi_t): t=1,\dots,n\}$ when constructing the {training and validation data. As small values of $Y_t < u_t$ are unlikely to impact the high $q$-quantile we seek to infer, we take the distribution of non-exceedances $F^\leq_t(\cdot)$ to be the empirical distribution of observations $\{{y}_t:{y}_t \leq u_t, t=1,\dots,n\}$, i.e., all observations that subceed the time-varying threshold $u_t$. We note that this distribution does not vary over $t$, but does vary with the prior draw of $\{u_t: t=1,\dots,n\}$.  }

The estimator is trained using early stopping \citep[see, e.g.,][]{prechelt2002early}; the validation loss is recorded at each epoch during training, which halts if the validation loss has not decreased for 10 epochs. We choose the optimal architecture for the neural networks, $\phi$ and $\boldsymbol{\psi}$ in Eq.~\eqref{eq:NE}, by minimising the risk for a test set of $1000$ parameter values (not used in training); this is provided in Table~\ref{tab:arch}. 
\begin{table*}[!hbt]
\caption{Optimal DNN architecture used in Section~\ref{sec:res:uni}. All layers used rectified linear unit (ReLU) activation functions, except the final layer, which used the identity function. }
\centering
\begin{tabular}{cccr} 
\toprule
\textbf{Function}  & \textbf{input dimension} & \textbf{output dimension}  & \textbf{parameters}\\
\midrule
\multirow{2}{*}{$\boldsymbol{\psi}$} & [1] & [48] & $49$ \\
 & [48] & [48]  & 2352 \\
 \hline
 $\mathbf{a}$ &[48] & [48] & 0 \\
 \hline
\multirow{2}{*}{$\phi$}  & [48] & [48]  & 2352 \\
& [48] & [1]  & 49 \\
\midrule
\multicolumn{2}{l}{total trainable parameters:} & \multicolumn{2}{r}{4802}\\
\bottomrule
\end{tabular}\label{tab:arch}
\end{table*}
NBEs are trained on GPUs randomly selected from within KAUST's Ibex cluster, see \url{https://www.hpc.kaust.edu.sa/ibex/gpu_nodes} for details (last accessed 13/07/2023).

\begin{figure}[!htb]
\centering
\begin{minipage}{0.45\linewidth}
\centering
\includegraphics[width=\linewidth]{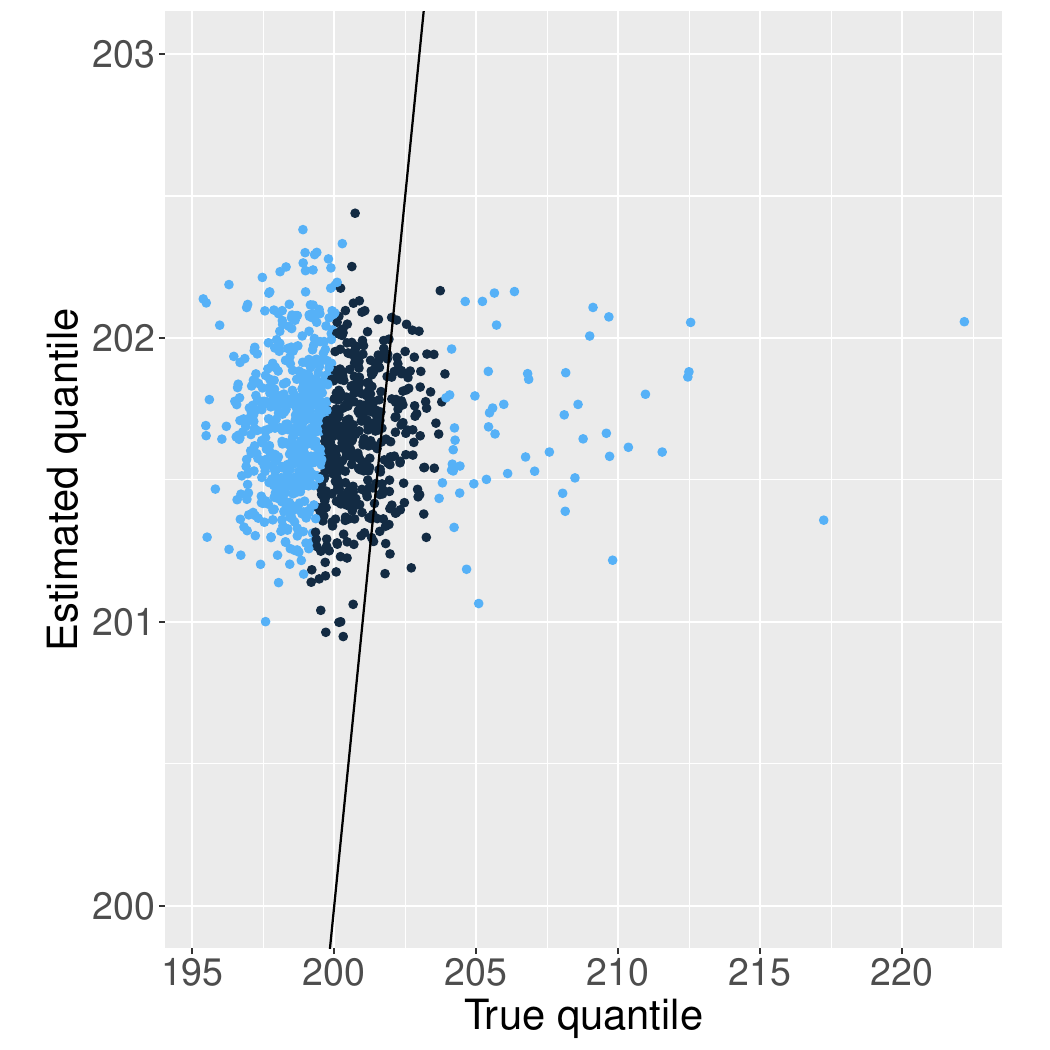} 
\end{minipage}
\begin{minipage}{0.45\linewidth}
\centering
\includegraphics[width=\linewidth]{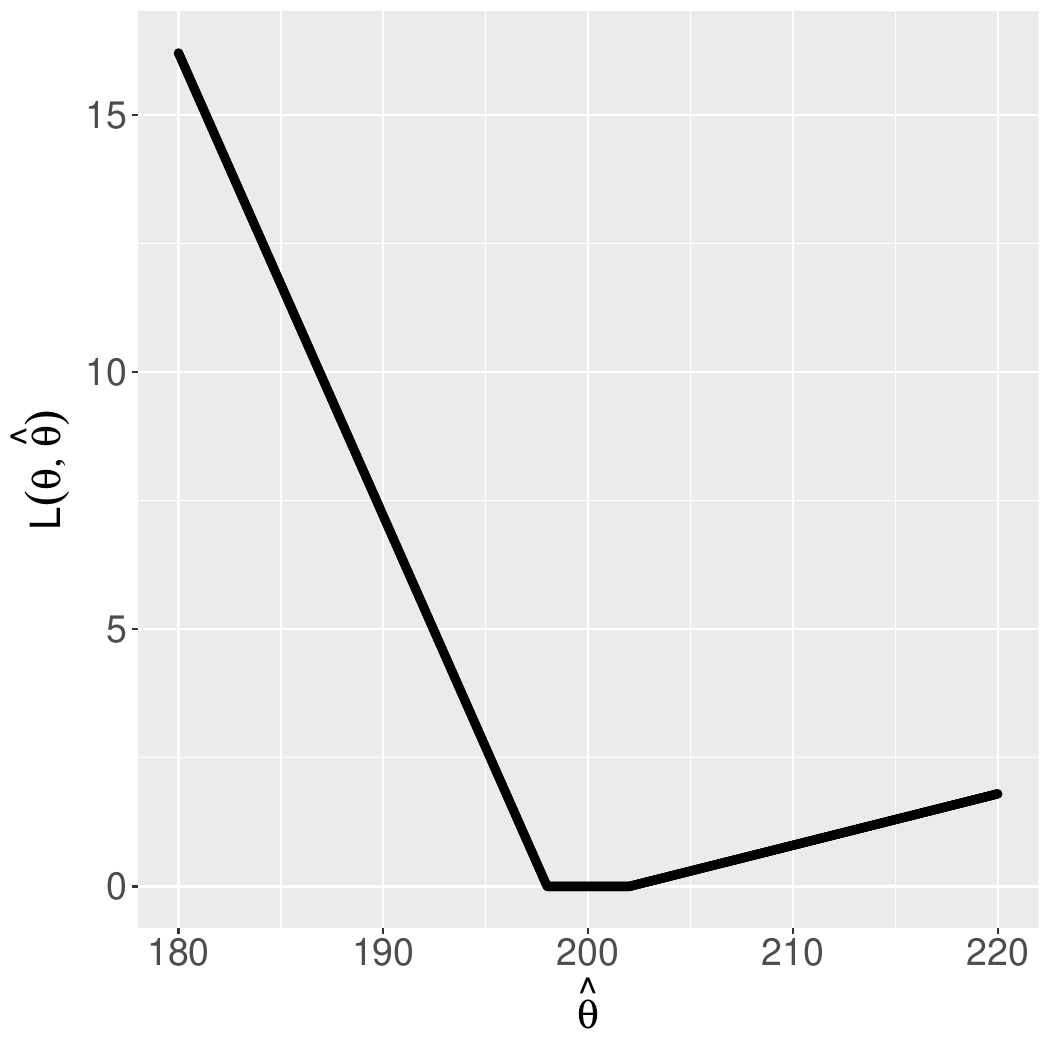} 
\end{minipage}
\caption{Left: Extreme quantile estimates for 1000 test data sets. Dark blue points denote those for which $L(\theta,\hat{\theta})=0$. Right: Loss function, $L(\theta_0,\hat{\theta})$ in Eq.~\eqref{eq:asym_loss}, for $\theta_0:=200$.}
\label{fig:NBE_res}
\end{figure}

To illustrate the efficacy of our estimator, Figure~\ref{fig:NBE_res} presents extreme quantile estimates for $1000$ test data sets. We observe that estimates are generally aligned to the left of the diagonal, i.e., the majority of estimates in Figure~\ref{fig:NBE_res} are overestimates. This is a consequence of the asymmetric loss function (also presented in Figure~\ref{fig:NBE_res}), which favours conservative estimates of the quantile. We estimate the $q$-quantile for $q=1-(6\times 10^4)^{-1}$ to be 201.25. An estimated 95\% confidence interval of (200.79, 201.73) was derived using a non-parametric bootstrap; due to the amortised nature of our neural estimator, non-parametric bootstrap-based uncertainty estimation is extremely fast, as the estimator does not need to be retrained for every new sample.

%%%%%%%%%%%%%%%%%%%%%%%%%%%%%%%%%%%%%%%%%%%%%%%%%%%%%%%%%%%%
%%% --- New Subsection --- %%%%%% --- New Subsection --- %%%
%%%%%%%%%%%%%%%%%%%%%%%%%%%%%%%%%%%%%%%%%%%%%%%%%%%%%%%%%%%%
\subsection{Extremal dependence estimation}
\label{sec:result-dep}
%%%%%%%%%%%%%%%%%%%%%%%%%%%%%%%%%%%%%%%%%%%%%%%%%%%%%%%%%%%%
%%% --- New Subsection --- %%%%%% --- New Subsection --- %%%
%%%%%%%%%%%%%%%%%%%%%%%%%%%%%%%%%%%%%%%%%%%%%%%%%%%%%%%%%%%%

For sub-challenge C3, we estimate $p_1$ and $p_2$ in Eq.~\eqref{Eq:C3} using the conditional extremes model described in Section~\ref{sec:CEM}. We perform model selection to identify the best covariates to include in the linear model for the dependence parameters, $\boldsymbol{\alpha}_{-i,t}$ and $\boldsymbol{\beta}_{-i,t}$ in Eq.~\eqref{eq:C3-paramform}, by minimising the AIC over the three separate model fits, i.e., $\mathbf{Y}_{-i,t} \mid (Y_{i,t} > u), i=1,2,3$, with $u$ set to the $90\%$ standard Laplace quantile. The best fitting model did not include any transformation (e.g, $\log$ or square) of the covariates, and uses a homogeneous $\boldsymbol{\beta}_{-i,t}$ function, i.e., with $\boldsymbol{\beta}_{-i,t}=\boldsymbol{\beta}_{-i}$ for all $t=1,\dots,n,$ and for $i=1,2,3$.

\begin{figure}[!htb]
\centering
\begin{minipage}{0.49\linewidth}
\centering
\includegraphics[width=\linewidth]{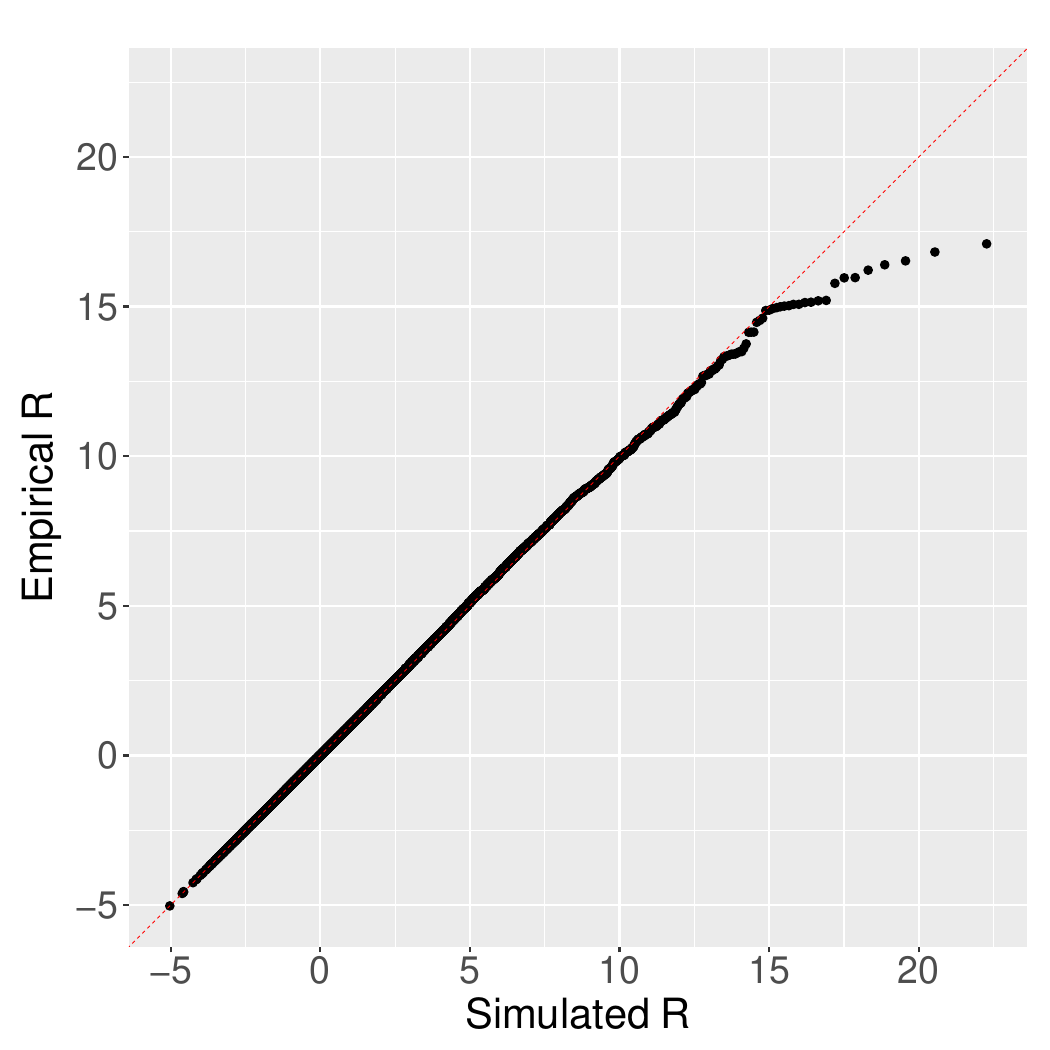} 
\end{minipage}
\begin{minipage}{0.49\linewidth}
\centering
\includegraphics[width=\linewidth]{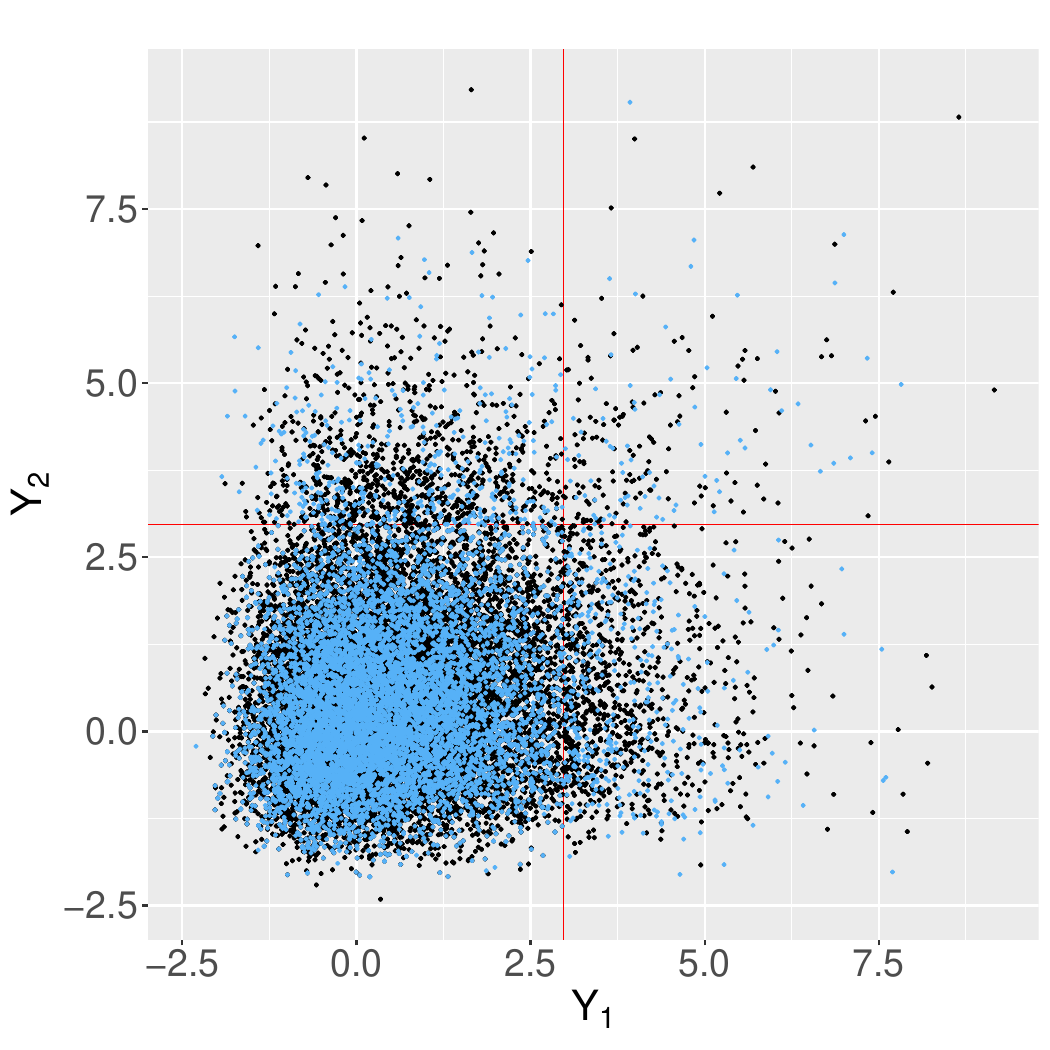} 
\end{minipage}
\begin{minipage}{0.49\linewidth}
\centering
\includegraphics[width=\linewidth]{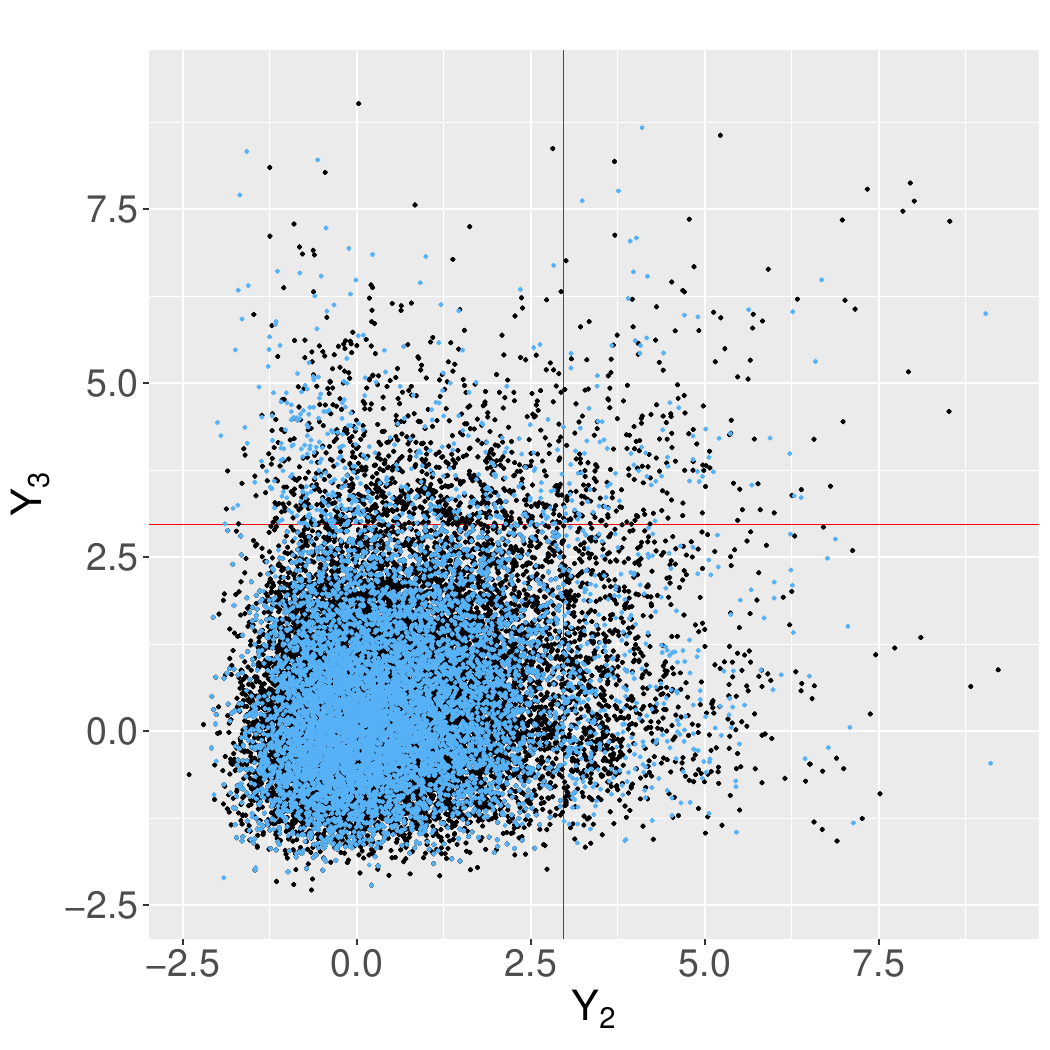} 
\end{minipage}
\begin{minipage}{0.49\linewidth}
\centering
\includegraphics[width=\linewidth]{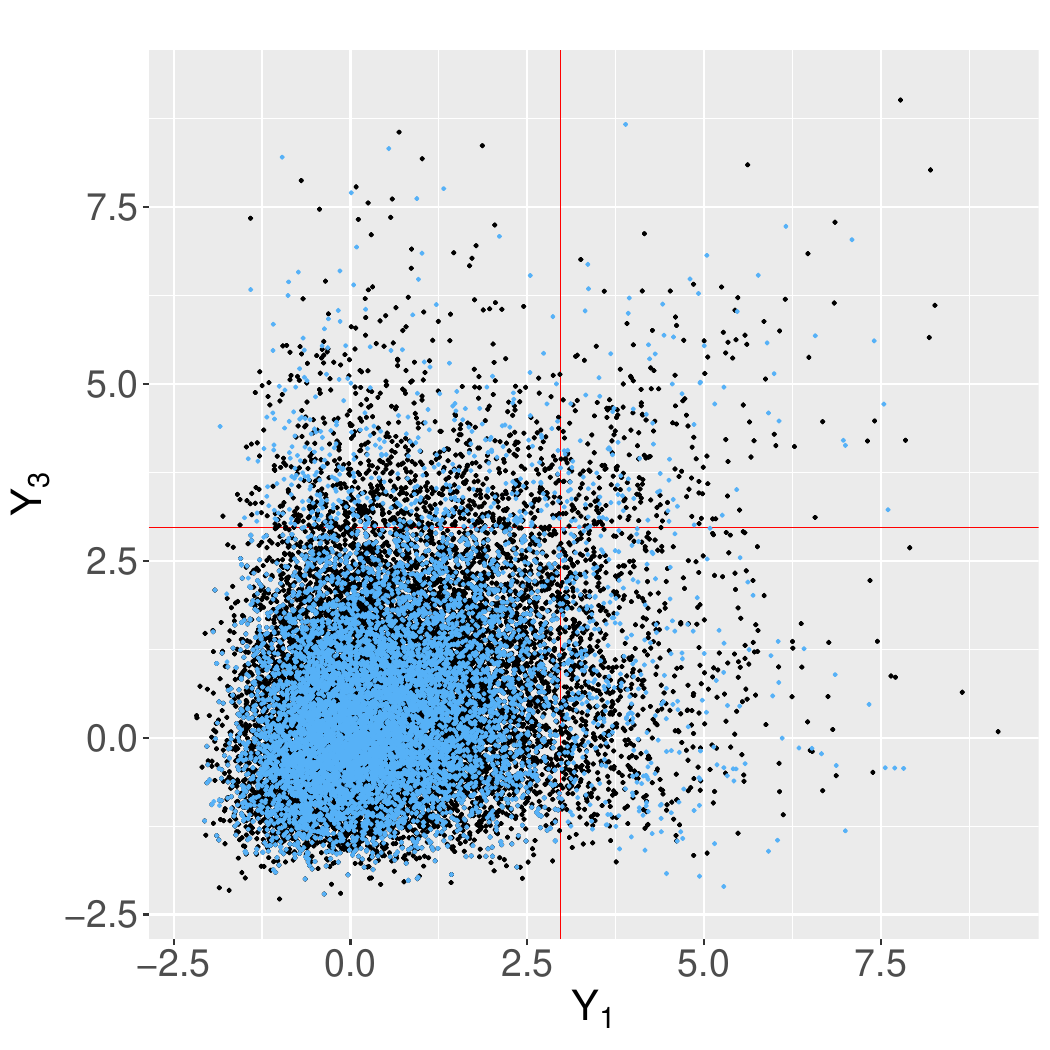} 
\end{minipage}
\caption{Top-left panel: Q-Q plot for the simulated and empirical sum $R=Y_{1}+Y_{2}+Y_{3}$. Other panels provide scatter plots of (black) observations and (blue) realisations from the fitted conditional extremal dependence models. Red vertical and horizontal lines correspond to the exceedance threshold $u$, i.e., the $95\%$ marginal quantile. }
\label{fig:C3}
\end{figure}

We use a visual diagnostic to optimise the exceedance threshold $u$. We fit the three conditional models for a grid of $u$ values, use Algorithm~\ref{sim-algo:2} {(with $N'=5N$)} to simulate $N:=10^7$ realisations  from the fitted model for $\mathbf{Y}=(Y_1,Y_2,Y_3),$ i.e., with time $t$ marginalised out, and then compute realisations of $R:=Y_{1}+Y_{2}+Y_{3}$. The upper tails of the aggregate variable $R$ are heavily influenced by the extremal dependence structure in the underlying $(Y_{1},Y_{2},Y_{3})$ \citep{richards2022tail}. Hence, we can exploit the empirical distribution of the aggregate as a diagnostic measure for the quality of the dependence model fit. {We choose the optimal $u$ as that which provides the best estimates for extreme quantiles of $R$; this is achieved through visual inspection of a Q-Q plot \citep[see, also,][]{richards2022modelling,richards2023joint}. For brevity, we test only two values of $u$: the $90\%$ and $95\%$ standard Laplace quantiles. We found the optimal $u$ to be the latter}, and we illustrate the corresponding Q-Q plot in Figure~\ref{fig:C3}. Table~\ref{table-4.3} provides the corresponding parameters estimates ${\hat{\boldsymbol{\alpha}}^{(0)}_{-i}}, {\hat{\boldsymbol{\alpha}}^{(1)}_{-i}},$ and $\hat{\boldsymbol{\beta}}_{-i}$, for $i=1,2,3$.

\begin{table*}[!t]
\caption{Sub-challenge C3: Parameter estimates  ${\hat{\boldsymbol{\alpha}}^{(0)}_{-i}}, {\hat{\boldsymbol{\alpha}}^{(1)}_{-i}},$ and $\hat{\boldsymbol{\beta}}_{-i}$, for the conditional extremal dependence model, $\mathbf{Y}_{-i,t} \mid Y_{i,t}>u$, $i=1,2,3$. Here, $\hat{\boldsymbol{\alpha}}^{(1,j)}_{-i}$ denotes the $j^{\text{th}}$ column of $\hat{\boldsymbol{\alpha}}^{(1)}_{-i}$ with $j=1$ and $j=2$ corresponding to the linear coefficient vector for \textit{Season} and \textit{Atmosphere}, respectively. Values in the $j^{\text{th}}$ row give the component of the parameter vector for $\mathbf{Y}_{-i,t}$  that corresponds to $Y_{j,t}$, $j=1,2,3$ (and so are missing for $j=i$). }
\centering
\small
\setlength\tabcolsep{4pt}
\begin{tabular}{l|cccc|cccc|cccc} 
\toprule
& \multicolumn{4}{c|}{$i=1$}& \multicolumn{4}{c|}{$i=2$}& \multicolumn{4}{c}{$i=3$}\\
&$\hat{\boldsymbol{\alpha}}^{(0)}_{-i}$&$\hat{\boldsymbol{\alpha}}^{(1,1)}_{-i}$&$\hat{\boldsymbol{\alpha}}^{(1,2)}_{-i}$&$\hat{\boldsymbol{\beta}}_{-i}$&$\hat{\boldsymbol{\alpha}}^{(0)}_{-i}$&$\hat{\boldsymbol{\alpha}}^{(1,1)}_{-i}$&$\hat{\boldsymbol{\alpha}}^{(1,2)}_{-i}$&$\hat{\boldsymbol{\beta}}_{-i}$&$\hat{\boldsymbol{\alpha}}^{(0)}_{-i}$&$\hat{\boldsymbol{\alpha}}^{(1,1)}_{-i}$&$\hat{\boldsymbol{\alpha}}^{(1,2)}_{-i}$&$\hat{\boldsymbol{\beta}}_{-i}$\\
\midrule
$Y_{1,t}$ &-&-&-&-&$-0.08$&$0.22$&$0.08$&$0.81$&$0.15$&$-0.33$&$0.01$&$0.97$\\
$Y_{2,t}$ &$0.56$&$-0.06$&$0.07$&$0.70$&-&-&-&-&$0.01$&$-0.07$&$0.01$&$0.45$ \\
$Y_{3,t}$ &$0.28$&$-0.02$&$0.07$&$0.82$&$0.24$&$-0.14$&$0.14$& $0.09$ &-&-&-&-\\
\bottomrule
\end{tabular}
\label{table-4.3}
\end{table*}

Example realisations from the fitted model are illustrated in Figure~\ref{fig:C3}. Good model fit is illustrated, as the realisations mimic the extremal characteristics of the observed data. Probabilities $p_1$ and $p_2$ are estimated empirically from the aforementioned ${N=10^7}$realisations of $(Y_1,Y_2,Y_3)$, and found to be $\hat{p}_1=3.13\times 10^{-5}$ and $\hat{p}_2=2.29\times 10^{-5}$, respectively. 

For sub-challenge C4, we estimate $p_3$ and $p_4$ in Eq.~\eqref{Eq:C4} using the non-parametric tail probability estimators defined in Eq.~\eqref{eq:pq4} and Eq.~\eqref{eq:pq5}, respectively. However, we first seek to decompose the high-dimensional random vector $\mathbf{Y}$ into strongly tail-dependent sub-vectors. Our reasoning is threefold: i) the estimator in Eq.~\eqref{eq:pq5} requires strong assumptions about the joint upper tail decay of the random vector $\mathbf{Y}$ which are unlikely to hold for high dimension $d$; ii) reliable estimation of the intermediate exceedance probability, $\widehat{p(\phi_*)}$ in Eq.~\eqref{eq:pq4}, becomes difficult as $d$ grows large; iii) \cite{krupskii2019nonparametric} observed through simulations that the estimator in Eq.~\eqref{eq:pq4} provides more accurate estimates when the true value of $\eta$ is close to one, i.e., when $\mathbf{Y}$ exhibits strong upper tail dependence.

\begin{figure}[t]
\centering
\begin{minipage}{0.6\linewidth}
\centering
\includegraphics[width=\linewidth]{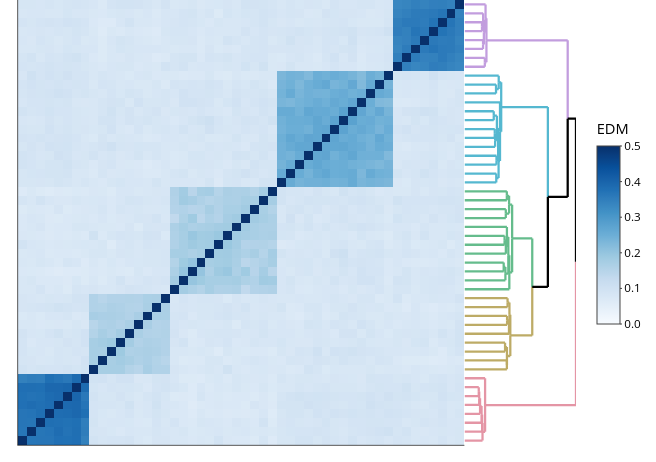} 
\end{minipage}
\begin{minipage}{0.39\linewidth}
\centering
\includegraphics[width=\linewidth]{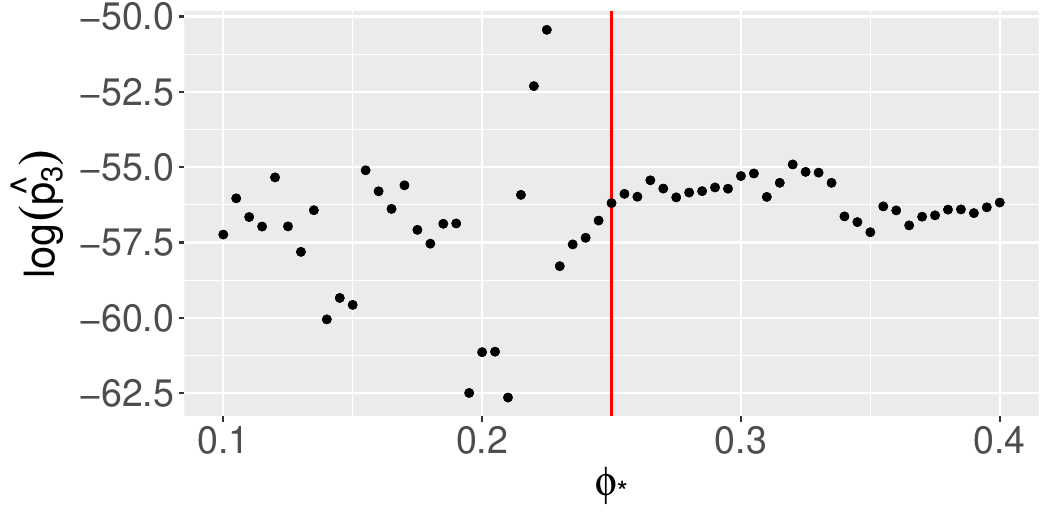} 
\includegraphics[width=\linewidth]{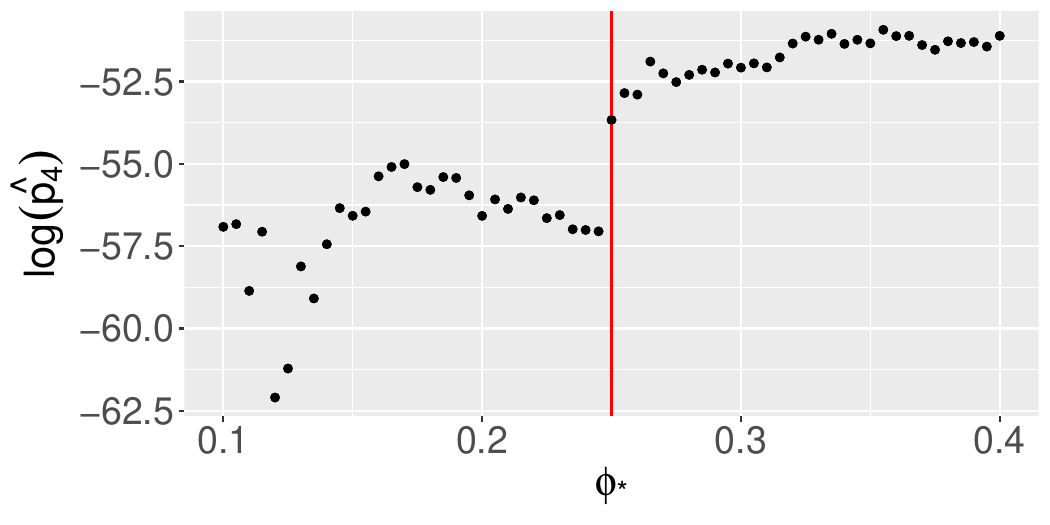} 

\end{minipage}
\caption{Left: Heatmap of pairwise EDM estimates grouped using hierarchical clustering, alongside the corresponding dendrogram. Right: Threshold stability plots of estimated log-exceedance probabilities, $\log(\hat{p_3})$ and $\log(\hat{p_4}),$ against the intermediate quantile $\phi_*$ used in their estimation. Red horizontal lines denote the optimal choice of $\phi_*$.}
\label{fig:C4}
\end{figure}

We identify sub-vectors by estimating the EDM between all pairs of components of $\mathbf{Y}$, with $\alpha=2$, $\|\cdot\|$ set to the $L_2$ norm, and with $u$ in Eq.~\eqref{EmpEDM} taken to be the empirical 0.99-quantile. Estimates are presented in a heatmap in Figure~\ref{fig:C4}, with values grouped using hierarchical clustering. An appropriate grouping can be found by visual inspection of Figure~\ref{fig:C4}; the clustered heatmap produces a block-diagonal matrix with five blocks (with size ranging from 8--13 components), and with elements of the off-diagonal blocks close to zero. For less well-behaved data, the number of clusters (and sub-vectors of $\mathbf{Y}$) can instead be determined by placing a cut-off at an appropriate point on the clustering dendrogram. 

We decompose $\mathbf{Y}$ into the five sub-vectors suggested by Figure~\ref{fig:C4}, which we denote by $\mathcal{Y}_i$ for  $i=1,\dots,5$. Estimation of $p_3$ then follows under the approximation
\begin{equation}
\label{eq:results_p3}
p_3 =\Pr(Y_1 >s_1, \dots, Y_{50} >s_1)\approx\prod_{i=1}^5\Pr\left(\bigcap\limits_{Y\in\mathcal{Y}_i} Y > s_1 \right),
\end{equation}
and similarly for $p_4$ (with the exceedance value $s_1$ or $s_2$ chosen appropriately). That is, we assume complete independence between the sub-vectors of $\mathbf{Y}$ to approximate $p_3$ and $p_4$. Exceedance probabilities for each sub-vector are estimated using the methodology described in Section~\ref{sec:nonpar}. We fix the intermediate quantile level $\phi_*$ in Eq.~\eqref{eq:pq4} across all sub-vectors and for estimation of both $p_3$ and $p_4$. An optimal value $\hat{\phi}_*$ is chosen via a threshold stability plot \citep[see, e.g.,][]{coles2001introduction}. In Figure~\ref{fig:C4}, we present estimates of $\log(p_3)$ and $\log(p_4)$ for a sequence of $\phi_*$ values in $[0.1,0.4]$. We choose $\hat{\phi}_*$ as the smallest $\phi_*$ such that there is visual stability in Figure~\ref{fig:C4} for estimates of both $\log(p_3)$ and $\log(p_4)$ when $\phi_* >\hat{\phi}_*$. In this case, we adopt $\hat{\phi}_*=0.25$ as the optimal value. We use a non-parametric bootstrap with 200 samples to assess uncertainty in estimates of $p_3$ and $p_4$. The resulting bootstrap median\footnote{We submitted the median over bootstrap estimates for the data competition.} estimates of $\log(p_3)$ and $\log(p_4$) (and their estimated 95\% confidence intervals) are $-59.38$ $(-68.46, -49.82)$ and $-55.24$ $(-60.12,-49.79)$, respectively.

%%%%%%%%%%%%%%%%%%%%%%%%%%%%%%%%%%%%%%%%%%%%%%%%%%%%%%%%%%%%
%%% --- End of Section --- %%%%%% --- End of Section --- %%%
%%%%%%%%%%%%%%%%%%%%%%%%%%%%%%%%%%%%%%%%%%%%%%%%%%%%%%%%%%%%
%%%%%%%%%%%%%%%%%%%%%%%%%%%%%%%%%%%%%%%%%%%%%%%%%%%%%%%%%%%%
%%% --- End of Section --- %%%%%% --- End of Section --- %%%
%%%%%%%%%%%%%%%%%%%%%%%%%%%%%%%%%%%%%%%%%%%%%%%%%%%%%%%%%%%%

%%%%%%%%%%%%%%%%%%%%%%%%%%%%%%%%%%%%%%%%%%%%%%%%%%%%%%
%%% --- New Section --- %%%%%% --- New Section --- %%%
%%%%%%%%%%%%%%%%%%%%%%%%%%%%%%%%%%%%%%%%%%%%%%%%%%%%%%
\section{Conclusion}\label{Sec:conclusion}   % --- %%%
%%%%%%%%%%%%%%%%%%%%%%%%%%%%%%%%%%%%%%%%%%%%%%%%%%%%%%
%%% --- New Section --- %%%%%% --- New Section --- %%%
%%%%%%%%%%%%%%%%%%%%%%%%%%%%%%%%%%%%%%%%%%%%%%%%%%%%%%

We proposed four frameworks for the estimation of exceedance probabilities and levels associated with extreme events. To estimate extreme conditional quantiles, we adopted an additive model that represents parameters in a peaks-over-threshold model as splines.  To estimate univariate quantiles in an unconditional setting, we constructed a neural Bayes estimator that estimates a quantile subject to a conservative loss function. Our new approach to quantile estimation is amortised, likelihood-free, and requires few parametric assumptions about the underlying distribution. For multivariate data, we considered two frameworks: i) in the presence of covariates, we adopted a non-stationary conditional extremal dependence model to capture linear trends in extremal dependence parameters, and ii) in the absence of covariates, we proposed a non-parametric estimator for multivariate tail probabilities that can be applied to high-dimensional ($d=50)$ data via a sparse decomposition using pairwise extremal dependence measures. We validated these modelling approaches {by using them to win} the EVA (2023) Conference Data Challenge \citep{editorial}.

{We illustrated the efficacy of additive GPD regression models for estimating extreme conditional quantiles. Machine learning-based extreme quantile regression models using, for example,  random forests \citep{gnecco2022extremal}, gradient boosting \citep{velthoen2023gradient}, and neural networks \citep{pasche2022neural, richards2023insights, richards2022regression} may offer a potentially more flexible alternative to additive models. }

We designed a neural Bayes estimator (NBE) to perform simulation-based inference for extreme quantiles. To train this estimator, we constructed an empirical prior measure $\pi(\cdot)$ for the extreme quantile, by simulating from a family of pre-defined peaks-over-threshold models. The resulting NBE was optimal with respect to the prior measure $\pi(\cdot)$.  Whilst our estimator performed well in the application, the underlying truth was a peaks-over-threshold model \citep{editorial}, and, hence, the truth distribution was contained within the class of distributions covered by $\pi(\cdot)$. We note that our estimator may not perform as well when the converse holds, and the true distribution of $Y$ is not approximated well by any family in $\pi(\cdot)$. Further study of the optimality properties of our proposed NBE is required and, in particular, its efficacy under model misspecification. 

We proposed an extension of the conditional extremal dependence regression model \citep{Winter2016} and highlighted its efficacy for estimation of extreme exceedance probabilities. Our conditional model performed well in practice, but alternative multivariate extremal dependence regression models, which rely on more classical assumptions about the extremal behaviour of $\mathbf{Y}$, e.g., regular or hidden regular variation, could have been tested \citep[see, e.g., ][]{cooley2012approximating, de2022regression, murphy2022modelling}. {In our application, we tested only two values for the exceedance threshold $u$ in \eqref{Eq:heff}, and quantified goodness-of-fit through visual inspection of Q-Q plots for the aggregate $R$ (see Figure~\ref{fig:C3}). A more rigorous testing procedure could be employed, whereby a grid of $u$ values are tested and the optimal $u$ is taken to minimise some numerical measure of fit; see, e.g., \cite{murphy2023automated}, who proposed an automated threshold selection procedure for peaks-over-threshold models.  }

{Finally, we proposed an extension of the non-parametric tail probability estimator of \cite{krupskii2019nonparametric} to account for different levels of marginal decay for components of a random vector. Whilst we were able to showcase the efficacy of our estimator by using it to win sub-challenge C4 of the EVA (2023) Conference Data Challenge, we have not provided any theoretical guarantees for the estimator; we leave this as a future endeavour.   }

%%%%%%%%%%%%%%%%%%%%%%%%%%%%%%%%%%%%%%%%%%%%%%%%%%%%%%%%%%%%
%%% --- End of Section --- %%%%%% --- End of Section --- %%%
%%%%%%%%%%%%%%%%%%%%%%%%%%%%%%%%%%%%%%%%%%%%%%%%%%%%%%%%%%%%
%%%%%%%%%%%%%%%%%%%%%%%%%%%%%%%%%%%%%%%%%%%%%%%%%%%%%%%%%%%%
%%% --- End of Section --- %%%%%% --- End of Section --- %%%
%%%%%%%%%%%%%%%%%%%%%%%%%%%%%%%%%%%%%%%%%%%%%%%%%%%%%%%%%%%%

%%%%%%%%%%%%%%%%%%%%%%%%%%%%%%%%%%%%%%%%%%%%%%%%%%%%%%
%%% --- New Section --- %%%%%% --- New Section --- %%%
%%%%%%%%%%%%%%%%%%%%%%%%%%%%%%%%%%%%%%%%%%%%%%%%%%%%%%
\section*{Declarations} %%%%%% --- New Section --- %%%
%%%%%%%%%%%%%%%%%%%%%%%%%%%%%%%%%%%%%%%%%%%%%%%%%%%%%%
%%% --- New Section --- %%%%%% --- New Section --- %%%
%%%%%%%%%%%%%%%%%%%%%%%%%%%%%%%%%%%%%%%%%%%%%%%%%%%%%%

\subsection*{Funding}
All authors were supported by funding from the King Abdullah University of Science and Technology (KAUST) Office of Sponsored Research (OSR) under Award No.\ {OSR-CRG2020-4394}.

\subsection*{Acknowledgments}
 The authors are grateful to Rapha\"el Huser and Matthew Sainsbury-Dale for helpful discussions and to Jonathan Tawn, Christian Rohrbeck, and Emma Simpson of Lancaster University, University of Bath, and University College London, respectively, for organisation of the data challenge that motivated this work. Support from the KAUST Supercomputing Laboratory is gratefully acknowledged. 

\bibliographystyle{apalike}
%%%%%%%%%%%%%%%%%%%%%%%%%%%%%%%%%%%%%%%%%%%%%%%%%%%%%%
%%% --- New Section --- %%%%%% --- New Section --- %%%
%%%%%%%%%%%%%%%%%%%%%%%%%%%%%%%%%%%%%%%%%%%%%%%%%%%%%%
\bibliography{References} %%%% --- New Section --- %%%
%%%%%%%%%%%%%%%%%%%%%%%%%%%%%%%%%%%%%%%%%%%%%%%%%%%%%%
%%% --- New Section --- %%%%%% --- New Section --- %%%
%%%%%%%%%%%%%%%%%%%%%%%%%%%%%%%%%%%%%%%%%%%%%%%%%%%%%%

\end{document}